\documentstyle[preprint,aps]{revtex} 
\preprint{
\vbox{
\hbox{TAUP-2408-97}
\hbox{TUM-T39-97-8}
}}

\input epsf 
\tighten
\newcommand{\llongrightarrow}{\relbar\joinrel\relbar\joinrel\longrightarrow}

\newcommand{\ppp}[1]{%
        \setbox0=\hbox{#1}%
        \kern-.02em\copy0\kern-\wd0
        \kern+.04em\copy0\kern-\wd0
        \kern-.02em\raise.0217em\box0}
\newcommand{\vek}[1]{
         \mathchoice{\mbox{\boldmath$#1$}}%
        {\mbox{\boldmath$#1$}}%
        {\ppp{$\scriptstyle#1$}}%
        {\ppp{$\scriptscriptstyle#1$}}}

\newcommand{\lsim}{$\raisebox{-0.8ex} {$\stackrel{\textstyle <}{\sim}$}$}
\newcommand{\gsim}{$\raisebox{-0.8ex} {$\stackrel{\textstyle >}{\sim}$}$}

\begin{document}  

\title{Coherent Photo- and Leptoproduction of 
       Vector Mesons from  Deuterium}

\author{L.~Frankfurt$^{a,e}$, W.~Koepf$^{b}$, 
J.~Mutzbauer$^{c}$, \\ G.~Piller$^{c}$, M.~Sargsian$^{c,f}$, 
M.~Strikman$^{d,e}$}
\date{\today{}}
\maketitle

\begin{center}

${(a)}$ School of Physics and Astronomy, Tel Aviv University,
Tel Aviv,  69978 Israel
 \\
${(b)}$ Physics Department, Ohio State University, Columbus,
OH 43210, USA 
\\
${(c)}$ Physik Department, Technische Universit\"{a}t M\"{u}nchen,
D-85747 Garching, Germany 
\\
${(d)}$ Pennsylvania State University, University Park,
     PA 16802, USA
\\
$(e)$ Institute for Nuclear Physics, St. Petersburg, Russia
\\
${(f)}$ Yerevan Physics Institute, Yerevan 375036, Armenia
\end{center}

\vspace*{2cm}

\begin{abstract}
We discuss the coherent  photo- and leptoproduction 
of vector mesons from deuterium at intermediate (virtual) photon 
energies, $3 \,GeV \,\lsim \,\nu\,\lsim \,30\,GeV$.   
These processes provide  several options to explore 
the space-time evolution of small size quark-gluon 
configurations.  
Furthermore, we study the dependence of the  production cross 
section on the energy and momentum transfer $t$ 
due to variations of the finite longitudinal interaction length. 
Kinematic regions are determined where the production cross 
section is most sensitive to the final state interaction  
of the initially produced hadronic  wave packet.
For unpolarized deuteron targets this double scattering 
contribution can be investigated 
mainly at large values of the momentum transfer $t$. 
For polarized targets kinematic windows sensitive to 
double scattering are available also at moderate $t$.
We suggest several methods for an investigation of color coherence 
effects at intermediate energies.
\end{abstract}

%%%%%%%%%%%%%
\newpage
%%%%%%%%%%%

\section{Introduction}

High-energy exclusive production processes from nucleon targets 
are determined by the transition of initial partonic wave 
functions  to final hadronic states. 
Interesting details about the corresponding amplitudes 
can be obtained by embedding the production process into nuclei,  
where the formation of a particular hadron is probed 
via the interaction with spectator nucleons 
(for recent reviews see e.g. \cite{CT}).

In this context  we discuss the coherent photo- and leptoproduction 
of vector mesons from unpolarized as well as polarized deuterium at 
photon energies $\nu \gsim 3\,GeV$ 
($q^{\mu}=(\nu,\vek q)$ is the photon four-momentum and $Q^2 = -q^2$):
\begin{equation}
\gamma^{(*)} + \stackrel{(\rightarrow)}{d} \longrightarrow V + d.
\label{reaction}
\end{equation}
The corresponding amplitudes can be split into two pieces:
a single scattering term in which only one nucleon participates 
in the interaction, and a  double scattering contribution.  
In this double scattering term the (virtual) photon interacts 
with one of the nucleons 
inside the target and produces an intermediate hadronic state   
which subsequently re-scatters from the second nucleon 
before forming the final state vector meson. 

At small $Q^2 \lsim 1\,GeV^2$ and intermediate energies,  
$3\,GeV\lsim \nu \lsim 30\,GeV$, exclusive vector meson production 
{}from deuterium is well understood in terms of vector meson dominance. 
Here, as seen from the laboratory frame, 
the final state vector meson is formed  
prior to the interaction with  the target 
(for a summary and references see e.g. \cite{BSYP}). 
On the other hand  in the limit of large $Q^2 \gg 1\,GeV^2$ 
perturbative QCD calculations show that photon-nucleon scattering  
produces a small-sized color singlet quark-gluon wave 
packet (ejectile) instead of a soft vector meson \cite{BFGMS}.\footnote{   
Note that this decrease of the ``size'' of the virtual photon 
at large  $Q^2$ has been already suggested before the advent 
of QCD  (see e.g. Refs.\cite{BSYP,ChW70,Kogut72}).}
At large energies $\nu$ the re-scattering  of such an ejectile 
with the second nucleon  should differ substantially from the 
re-scattering of a soft vector meson. 
Therefore the magnitude of the double scattering contribution 
to the production process contains 
interesting informations about the initially produced ejectile and 
its evolution while propagating through the target. 

The ejectile wave packet and its propagation are  characterized 
by the following scales: 
the average {\bf transverse size} of the wave packet 
which for the case of longitudinal photons is
$b_{ej}\approx  4 \ldots 5/Q$ for the contribution 
of the minimal Fock space component at $Q^2 \gsim 5 GeV^2$ \cite{FKS}.
For these $Q^2$ it amounts to less than 
a third of the typical diameter of a $\rho$-meson ($\approx 1.2\,fm$).
Furthermore, the initially produced small quark-gluon wave packet does 
not, in general, represent an eigenstate of the strong interaction 
Hamiltonian.
Expanding the ejectile in hadronic eigenstates one 
finds  that  inside a nuclear target all hadronic components, except 
the measured vector meson, are filtered out via final state interaction 
after a typical {\bf formation time}:  
$\tau_f \approx 2\nu/\delta m_V^2$. 
Here $\delta m_V^2$ is the characteristic difference of the squared  
masses of low-lying vector meson states  which is related to
the inverse slope of the corresponding Regge trajectory 
($\delta m_V^2 \sim 1 GeV^2$).
If the formation time is larger than the nuclear radius
the coherence between the hadronic eigenstates, which describe 
the ejectile, is kept 
on the scale of the target dimensions. 
Then the transverse size of the ejectile is frozen during its 
penetration through the target. 
Contributions from re-scattering processes should therefore decrease 
with rising $Q^2$ at large photon energies. 
This phenomenon is commonly called color coherence or color transparency
\cite{CT}.  

However, it has been understood for more than twenty years that 
dominant contributions  to high-energy, photon-induced 
processes result  from large longitudinal space-time intervals 
which increase with energy \cite{GIP,Gribov70,LL}. 
As soon as they are of the order of the average nucleon-nucleon 
distance in nuclei  nuclear effects become important.
A well known experimental confirmation of this fact  
can be found in nuclear deep-inelastic scattering at 
small values of the Bjorken scaling variable $x = Q^2/2M\nu$,
where $M$ is the nucleon mass.  
Here the increase of the longitudinal interaction distance 
$\sim 1/2Mx$ (see e.g. \cite{Ioffe})
which dominates the reaction at small $x$ leads to the leading twist 
shadowing of  parton distributions in nuclei at $x \ll 0.1$  
(for a recent review see Ref.\cite{Arneodo}).

For diffractive vector meson production from nuclei 
the situation is similar. 
Here the {\bf characteristic longitudinal interaction 
length}\footnote{
A detailed  discussion of the $t$-dependence of $\lambda$ 
is presented in Sec.\ref{3b}.}
$\lambda \approx 2\nu/(m_V^2 + Q^2)$
is determined by the minimal momentum transfer, $t_{min} = -1/\lambda^2$, 
required for the diffractive production of the vector meson 
with invariant mass $m_V$. In the kinematic domain of intermediate 
photon energies  $\nu \lsim\,30 \,GeV$ and $Q^2 \gsim 1\,GeV^2$, which
is accessible for example 
at HERMES, $\lambda$ is of the order of typical nuclear dimensions 
and can have a major influence on the scattering process. 
Note that for $Q^2 > 3\,GeV^2$ the same is true even for large 
photon energies $\nu \lsim 150\,GeV$ as available at FNAL.
Therefore if one investigates the double scattering contribution 
to diffractive vector meson production in different kinematic regions  
one should be careful in interpreting a variation of the 
re-scattering strength  as a modification of the  ejectile wave function. 
First,  possible effects arising from a variation of the characteristic 
longitudinal interaction length have to be investigated. 
For total production cross sections  
and differential cross sections at small momentum transfers, $t\approx 0$,
they are accounted for in the framework of the 
Glauber multiple scattering theory \cite{BSYP,Spital76,Yennie71,HKN}. 
However, as we will show,  Glauber theory does not reproduce the  
proper longitudinal interaction length in vector meson production 
at $t\ne 0$, since it neglects nuclear recoil effects which 
turn out to be important for light nuclei.

The privileged use of deuterium as a target has several reasons: 
first the deuteron is the simplest bound state of nucleons 
and therefore the ``best'' understood nucleus. Its wave function
is known within a few percent accuracy at least up to 
internal momenta $\lsim 400\,MeV/c$, and
further insights on the high momentum tail of the 
deuteron wave function are expected from current experiments at 
MAMI, NIKHEF and TJNAF. 
This provides many possibilities to investigate 
double scattering processes on different characteristic 
length scales using for example polarized targets 
or tagging on specific break-up proton-neutron final states.

Note that in inclusive vector meson production from 
heavy nuclei 
the formation length is larger than the size of the nuclear target only 
at large photon energies. This makes an analysis of color 
coherence effects at moderate energies more difficult. 
Also many exclusive channels in vector meson production from 
heavy nuclei suffer from theoretical uncertainties related to 
the more complicated final state interaction of the detected hadrons.

This paper is organized as follows: in Sec.\ref{sec:amplitude} 
we review the derivation of the single and double scattering amplitudes
for diffractive vector meson production. 
The differential production cross section for coherent  
processes is derived in Sec.\ref{sec:coherent}.
There, after showing evidence for double scattering events, we discuss 
effects due to the kinematic dependence 
of the finite longitudinal interaction length 
and suggest corresponding experiments feasible at TJNAF and HERMES.
We then propose  signatures for color coherence at large 
and moderate momentum transfers suitable for investigations at 
HERMES. Conclusions follow in Sec.\ref{sec:conclusions}

\section{Photon-deuteron scattering amplitude}
\label{sec:amplitude}

In view of the high energy of the incident photon 
we derive the vector meson production amplitude  within 
the eikonal approximation. 
This can be done for coherent 
as well as incoherent diffractive production processes. 
In the next section we then focus on the coherent case.
The following notations are used:  $p_d^{\mu} = (M_d,\bf 0)$ 
is the deuteron four-momentum in the laboratory frame  which we will use
throughout this investigation.
The final state proton-neutron system with invariant mass $M_f$  
is either a deuteron or a  proton-neutron continuum state ($pn$).  
Its four-momentum is denoted by $p_d'$. 
Furthermore, $q^{\mu} = (\nu,{\bf 0}_{\perp},\sqrt{Q^2 + \nu^2})$ 
and $k_V$ are the momenta of the  
incident photon and the produced vector meson, respectively. 
They determine the momentum transfer to the deuteron,  
$l^{\mu} = (l_0,\vek l) = q^{\mu} - k_V^{\mu}$. 

In the impulse approximation the vector meson is 
produced either from  the proton or the neutron. 
The corresponding diagram is shown in Fig.1a.
In a non-relativistic treatment of the deuteron target and the 
final $pn$ system we obtain for the single scattering amplitude: 
\begin{eqnarray} \label{eq:ia}
F^{(a)} &=&  
f^{\gamma^*p\rightarrow Vp}(\vek l)\,
S^{jj'}_{df}\left(-{\vek l_{\perp}\over 2}, {l_{\_} \over 2}\right)~+~
f^{\gamma^*n\rightarrow Vn}(\vek l)\,
S^{jj'}_{df}\left({\vek l_{\perp}\over 2}, -{l_{\_} \over 2}\right). 
\end{eqnarray}
Here only the dependence of the photon-nucleon production 
amplitudes  $f^{\gamma^* N\rightarrow VN}$  on the 
corresponding three-momentum transfer is shown explicitly.  
Furthermore we  have introduced the non-relativistic transition form factor 
(the corresponding relativistic expression is given in Appendix A):
\begin{equation}
S^{jj'}_{df}(\vek l) = \int d^3 k\, 
\psi^{j}_d\left(\vek k+{\vek l \over 2}\right)
\psi^{j'\dagger }_f\!\left(\vek k-{\vek l\over 2}\right) = 
\int d^3r\, 
\psi^j_d(\vek r)\psi^{j'\dagger}_f(\vek r)e^{-i\vek l\cdot \vek r},
\label{eq:traFF}
\end{equation}
where $\psi_d^j$ is the wave function of the deuteron target with 
spin quantum numbers $j$, and ${\psi_f^j}'$ denotes the 
wave function of the final $pn$ system with spin $j'$. 
Bound state  wave functions are normalized according to 
$\int d^3k\,|\psi_{d}(\vek k)|^2 =1$, 
while two-nucleon continuum wave functions obey 
$\int d^3k\,\psi^{P}_f(\vek k)\psi^{P'+}_f(\vek k) = 
\delta^3(\vek P-\vek P')$, 
where $\vek P$ and $\vek P'$ are the 
momenta of the $pn$ state with energy $E=\vek P^2 /2M$.
Throughout the paper all amplitudes are normalized according to 
$Im \,f^{VN\rightarrow VN}(l=0) = \sigma_{VN}$, 
the latter being the total vector meson-nucleon cross section. 
Here and in the following we temporarily suppress the spin dependence of 
the scattering amplitudes.
Note that Eq.(\ref{eq:ia}) corresponds to the 
well known  result of Refs.\cite{FrGl66,bert} except 
for the presence of \
$l_{\_} = l_0 - l_z = \sqrt{M^2_{f}+\vek l^2}-M_d - l_z$   
in the form factor. The latter accounts for the recoil of the 
two-nucleon system (see Appendix A).

In the double scattering contribution the vector meson 
is produced via an intermediate hadronic state $h$ as shown in 
Fig.1b. 
The corresponding amplitude reads:
\begin{eqnarray} 
F^{(b)} &=&  -\int {d^3 p_s  d^3 p'_s\over  (2\pi)^3}\,
\psi^j_d\left( - \vek p_s\right)
{f^{\gamma^*p\rightarrow hp}(\vek l_{\perp}  - \vek k_{\perp}) 
\,k_h^0\, 
f^{hn\rightarrow Vn}(\vek k_{\perp})
\over D_h(q-l+ k)}
\nonumber \\ 
&& \hspace{5cm}
\times
\psi^{j' \dagger}_f\!
\left({\vek l_{\perp}\over 2}-\vek k_{\perp}-\vek p_{s\perp},
-{l_{\_}\over 2} + k_{\_}-p_{s z} \right)  
+(p\leftrightarrow n).
\label{eq:ds}
\end{eqnarray}
where $p_s$ and $p_s'$ are the four-momenta  of the spectator nucleon 
before and after the re-scattering of the intermediate 
hadronic state $h$ and $k=p_s'-p_s$.
Similar to single scattering  we account for the 
recoil of the two-nucleon final state (see Appendix A, Eq.(\ref{llc5})).
Note that  in the derivation of Eq.(\ref{eq:ds}) completeness over the 
two-nucleon intermediate states has been used. 

The intermediate state $h$  
carries an invariant mass $m_h$ and a four-momentum 
$k_h = k_V + k = q - l + k$.  
Its propagator leads to the denominator $D_h(k_h) = k_h^2 - m_h^2 + i\epsilon$.
In Eq.(\ref{eq:ds}) we have 
assumed  that the diffractive amplitudes $f$ depend on the 
corresponding transverse momentum transfer only. 
This is justified by the fact  that momenta
characteristic for bound nucleons have negligible effects  on the 
amplitudes $f$. Consequently 
the light-cone momentum $l_-$ enters effectively only 
in the form factor and propagator of the intermediate  hadronic state.
After  introducing  $P = (p'_s+p_s)/2$ 
one can factorize the  integrations in (\ref{eq:ds}) 
and  obtains:
\begin{eqnarray}
F^{(b)} &=& - \int {d^3 k \over  (2\pi)^3}\,
S_{df}^{j j'}\left(
-{\vek l_{\perp}\over 2} + \vek k_{\perp}, 
{l_{\_}\over 2}-k_{-} 
\right) 
{f^{\gamma^*p\rightarrow hp}({\vek l}_{\perp}-\vek k_{\perp}) 
\,k_h^0\,
f^{hn\rightarrow Vn}(\vek k_{\perp})\over 
 D_h(q -{l}+ k)} +(p\leftrightarrow n).   
\label{eq:pref}
\end{eqnarray} 
Substituting $ -l/2 + k \rightarrow  k$ leads to: 
\begin{eqnarray}
F^{(b)} &=& - \int {d^3 k \over  (2\pi)^3}\,
S_{df}^{j j'}\left(\vek k_{\perp}, -k_{-}\right) 
{f^{\gamma^*p\rightarrow hp}({\vek l_{\perp}\over 2}-\vek k_{\perp}) 
\,k_h^0\,
f^{hn\rightarrow Vn}({\vek l_{\perp}\over 2} + \vek k_{\perp})\over 
 D_h(q -{l\over 2}+ k)} +(p\leftrightarrow n).   
\label{eq:prefb}
\end{eqnarray} 
Within the non-relativistic approximation for the initial and final 
proton-neutron system 
the denominator of the intermediate hadron can be reduced to:
\begin{eqnarray}
D_h(q-{l\over 2}+k) &=& \left(q-{l\over 2}+k\right)^2 - m_h^2 + i\epsilon  
\nonumber \\
&\approx& 
2 k_h^0 \left[ k_{-} - {l_{\_}\over 2} -  
{({\vek l\over 2}-\vek k)^2\over 2 \nu} - 
{Q^2 + m_h^2\over 2 \nu} + i\epsilon\right],
\label{eq:den1}
\end{eqnarray}
where only leading terms with respect to the photon 
energy are kept. 
The denominator in Eq.(\ref{eq:den1}) can be simplified further 
using the empirical fact that  diffractive amplitudes 
can be parameterized by 
$f(\vek k) \sim e^{-{B\over 2} \vek k^2}$, 
with a typical slope of $B \sim (5-8)\,GeV^{-2}$ 
depending on $Q^2$ and $\nu$\cite{BSYP,slopes}. 
Consequently, at large photon energies we can  neglect 
the term $({\vek l\over 2}-\vek k)^2 /2 \nu \lsim 
{1}/{B\nu}$.
Furthermore we omit $k_0\approx \vek k^2_{\perp}/ 2M$ 
which is small due to the 
transition form factor in Eq.(\ref{eq:prefb}) 
which favors contributions from $\vek k_{\perp}^2 \approx 0$.
Substituting Eq.(\ref{eq:den1}) into Eq.(\ref{eq:prefb}) 
gives:
\begin{eqnarray}
F^{(b)} &=& - \int {d^3 k \over  2 (2\pi)^3}\,
S_{df}^{j j'}\left(\vek k \right) 
{f^{\gamma^*p\rightarrow hp}({\vek l_{\perp} \over 2 }-\vek k_{\perp}) 
f^{hn\rightarrow Vn}({\vek l_{\perp} \over 2 }+\vek k_{\perp})\over 
- k_z - \Delta_h + i \epsilon } +(p\leftrightarrow n), 
\label{eq:pref2}
\\
&& \mbox{with} 
\quad \Delta_h = \frac{l_{\_}}{2} + \frac{Q^2 + m_h^2}{2 \nu}
= \frac{Q^2 + 2m_h^2-m_{V}^2+t}{4 \nu}.
\label{Delta}
\end{eqnarray} 
Performing the $k_z$-integration leads to:
\begin{eqnarray}
F^{(b)} &=& 
{i\over 2\sqrt{2\pi}} 
\int{d^2 k_\perp  \over  (2\pi)^2} dz\,\Theta(-z)\,
S^{j j'}_{df}(\vek k_\perp,z)
\,e^{i\Delta_h z} \nonumber \\
& & \qquad\qquad \times f^{\gamma^*p\rightarrow hp}
\left({\vek l_{\perp}\over 2}-\vek k_{\perp}\right) 
f^{hn\rightarrow Vn}\left({\vek l_\perp\over 2}+\vek k_{\perp}\right) 
+~(p\leftrightarrow n),  
\label{eq:ds2}
\end{eqnarray}
where the transition form factor in the mixed representation is 
defined via:
\begin{equation}
S^{j j'}_{df}(\vek k_{\perp},k_z) = {1\over \sqrt{2\pi}}
\int dz\,S^{j j'}_{df}(\vek k_{\perp},z)e^{-ik_zz}. 
\label{eq:tranFFmixed}
\end{equation}
Adding explicitly the contribution where first the neutron and then 
the proton interacts gives:
\begin{equation} \label{eq:DS}
F^{(b)} = \tilde F^{(b)} + \Delta F^{(b)},  
\end{equation}
with: 
\begin{eqnarray}   
\tilde F^{(b)} &=& 
{i\over 2} 
\int {d^2 k_\perp  \over  (2\pi)^2}\,
S^{j j'}_{df}(\vek k_\perp,-\Delta_h) 
f^{\gamma^*p\rightarrow hp}
\left({\vek l_\perp\over 2}-\vek k_\perp\right) 
f^{hn\rightarrow Vn}
\left({\vek l_\perp\over 2}+\vek k_\perp\right),
\label{eq:ds_lead}
\\
\Delta F^{(b)} &=& {i\over 2\sqrt{2\pi}} 
\int {d^2 k_\perp  \over  (2\pi)^2} dz \,\Theta(z) \,
\times \\
& &\quad
\left[
S^{j j'}_{df}(-\vek k_\perp,z)e^{-i\Delta_h z} 
f^{\gamma^*n\rightarrow hn}
\left({\vek l_{\perp}\over 2}-\vek k_\perp\right) 
f^{hp\rightarrow Vp}
\left({\vek l_{\perp}\over 2}+\vek k_\perp \right)
\right. \nonumber \\ 
& &\qquad\left. -~
S^{j j'}_{df}(\vek k_\perp,z)e^{i\Delta_h z}
f^{\gamma^*p\rightarrow hp}
\left({\vek l_\perp\over 2}-\vek k_\perp\right) 
f^{hn\rightarrow Vn}
\left({\vek l_\perp \over 2}+\vek k_\perp\right)
\right]\nonumber .
\end{eqnarray}
Since $\Delta F^{(b)}$ is only a small correction to $\tilde F^{(b)}$ 
we neglect in the former the difference between the proton and 
neutron amplitudes. 
For specific target polarizations 
the form factor $S^{j j'}_{df}$ is invariant under rotations 
in the transverse plane, i.e. 
$S^{j j'}_{df}(-\vek k_\perp,z) = S^{j j'}_{df}(\vek k_\perp,z)$. 
Then one obtains:
\begin{equation}
\Delta F^{(b)}  =   -{1\over \sqrt{2\pi}} 
\int {d^2 k_\perp  \over  (2\pi)^2}\,
\Delta S^{j j'}_{df}(\vek k_\perp,-\Delta_h)
\,f^{\gamma^*N\rightarrow hN}
\left({\vek l_\perp\over 2}-\vek k_\perp\right) 
f^{hN\rightarrow VN}
\left({\vek l_{\perp}\over 2}+\vek k_{\perp}\right),
\label{eq:delta_f}
\end{equation}
with
\begin{equation}
\Delta S^{j j'}_{df}(\vek k_\perp,-\Delta_h) = 
\int dz \,\Theta(z) \,
S^{j j'}_{df}(\vek k_\perp,z) 
\,\sin\left(-\Delta_h z\right) \ .
\label{delta_s}
\end{equation}
It is important to note that the double scattering amplitude 
in Eq.(\ref{eq:DS}) explicitly includes  the 
energy and momentum dependence of the 
longitudinal photon interaction length. 
It therefore  differs form the corresponding amplitude derived for 
hadron-deuteron scattering processes \cite{FrGl69}. 
In addition, the derived single (\ref{eq:ia}) and 
double scattering amplitude (\ref{eq:DS})  account for
the recoil of the final $pn$ system which is not included  
in the conventional approach of Ref.\cite{FrGl69}.
The results of the latter can be obtained from 
Eqs.(\ref{eq:ia}) and (\ref{eq:DS}) 
after neglecting recoil effects by
taking the limit $\Delta_h \rightarrow 0$ and 
replacing $\gamma^* \rightarrow h = V$.

Collecting the results for single (\ref{eq:ia}) and double scattering  
(\ref{eq:DS})  and 
summing over all possible intermediate hadronic states $h$ 
gives for the vector meson production amplitude:  
\begin{eqnarray} \label{eq:amplitude} 
F_{df}^{j j'} &=& F^{(a)} + F^{(b)} =   
f^{\gamma^*N\rightarrow VN}(\vek l)
\left[
S^{j j'}_{df}\left(-{\vek l_{\perp}\over 2}, 
{l_{\_} \over 2}\right)~+~
S^{j j'}_{df}\left({\vek l_{\perp}\over 2}, 
-{l_{\_} \over 2}\right)\right] 
\nonumber \\ 
&+&
{i\over 2}\sum_h \int 
{d^2 k_\perp  \over  (2\pi)^2}
f^{\gamma^*N\rightarrow hN}
\left({\vek l_\perp\over 2}-\vek k_\perp\right) 
f^{hN\rightarrow VN} 
\left({\vek l_\perp\over 2}+\vek k_\perp\right) \nonumber \\
&&\hspace{1.5cm}\times
\left[
S^{j j'}_{df}(\vek k_\perp,-\Delta_h) + 
\frac{2 i}{\sqrt{2\pi}} 
\Delta S^{j j'}_{df}(\vek k_\perp,-\Delta_h)
\right].
\end{eqnarray}

To investigate the production process in different  kinematic domains 
it is useful to express the transition amplitude 
$f^{\gamma^* N \rightarrow h N}$ in a hadronic basis for 
the incoming virtual photon (see e.g. \cite{Gribov70,Spital76}):
\begin{equation}
f^{\gamma^*N\rightarrow hN} = 
\sum_{h'}{<0|\epsilon_{\gamma^*}\cdot J^{em}|h'> d\tau_{h'}
\over E_{h'}-\nu}f^{h'N\rightarrow h N}\ ,
\label{eq:spectral}
\end{equation}
where $\epsilon_{\gamma^*}$ is the polarization vector of the 
virtual photon and $J^{em}$ is the electromagnetic current. 
$E_{h'}$ denotes the energy of the intermediate hadronic state $h'$ and 
$d\tau_{h'}$ stands for the corresponding  phase-space factor.
The hadronic representation in (\ref{eq:spectral})  
is based on the existence of a spectral representation for the 
electromagnetic scattering 
operator \cite{Bjorken}. 
Inserting Eq.(\ref{eq:spectral}) 
into the amplitude (\ref{eq:amplitude}) leads to: 
\begin{eqnarray}
F_{df}^{j j'}  & = & 
\sum_{h'}{<0|\epsilon_{\gamma^*}\cdot J^{em}|h'> d\tau_{h'}
\over E_{h'}-\nu}
f^{h' N\rightarrow VN}(\vek l)
\left[
S^{j j'}_{df}\left(-{\vek l_{\perp}\over 2}, 
{l_{\_} \over 2}\right)~+~
S^{j j'}_{df}\left({\vek l_{\perp}\over 2}, 
-{l_{\_} \over 2}\right)
\right] 
\nonumber \\
& + &  {i\over 2}\sum\limits_{h',h} 
\int \frac{d^2k_{\perp}}{(2\pi^2)^2}{<0|\epsilon_{\gamma^*}\cdot J^{em}|h'> 
d\tau_{h'}\over E_{h'}-\nu}
f^{h'N\rightarrow hN}
\left({\vek l_\perp   \over 2}-\vek k_\perp\right) 
f^{hN\rightarrow VN} 
\left({ \vek l_\perp   \over 2}+\vek k_\perp\right) 
\nonumber \\
&& 
\times 
\left[
S^{j j'}_{df}(\vek k_\perp,-\Delta_h) + 
\frac{2 i}{\sqrt{2\pi}} 
\Delta S^{j j'}_{df}(\vek k_\perp,-\Delta_h)
\right].
\label{eq:ampl_ii}
\end{eqnarray}
Although the amplitude in Eq.(\ref{eq:ampl_ii}) 
cannot be calculated explicitly in a model independent way, 
it reveals three different energy domains of the production 
process.

The first region corresponds to intermediate photon energies and low 
momentum transfers $Q^2 \lsim 1~GeV^2$.
Here contributions to the scattering amplitude (\ref{eq:ampl_ii}) from 
intermediate states $h'$ with large invariant masses  $m_h'$ 
are  suppressed by large energy denominators:
\begin{equation} \label{eq:en_den}
E_{h'} - \nu  \approx  \frac{m_{h'}^2 + Q^2}{2 \nu}.
\end{equation}
The possible restriction to small mass intermediate states 
leads to the vector meson dominance (VMD) model 
(for a review see e.g. Ref.\cite{BSYP}). 

At high energies but fixed $Q^2$ contributions from  
intermediate hadronic states with large masses ($m_{h'}^2 \gg Q^2$)
are important. 
They are usually described through the triple reggeon 
formalism \cite{TRA}. 

The third kinematic domain corresponds to large momentum transfers,  
$Q^2\gg 1~GeV^2$, and small values of $x \ll 1/4 M R_A$, 
where $R_A$ is the radius of the target nucleus.
Here it is legitimate to use closure over  intermediate 
hadronic  states $h$ and $h'$  and substitute  in 
(\ref{eq:ampl_ii})  the sums over hadronic states by  
a quark-gluon wave packet which can be calculated 
within perturbative QCD. 
For example in the case  of longitudinally polarized photons at large $Q^2$
short distance dominance  leads  
to the interaction of a color-dipole, quark-antiquark pair 
with small transfer size \cite{BFGMS,FKS,AFS,FGKSS}.

We want to investigate the $Q^2$-dependence of the 
re-scattering amplitude  in the kinematic domain $0 <Q^2 < 10\,GeV^2$ 
and moderate photon energies, such that closure over intermediate 
hadronic states as discussed above is not applicable. 
To account for effects which result from the kinematic 
dependence  of the effective longitudinal interaction 
length we perform a baseline calculation within 
the framework of vector meson dominance. 
We then consider processes which are sensitive to 
re-scattering, but to a good approximation 
independent from the initial production process. 
Thus we obtain signatures for color coherence and 
avoid uncertainties which result  from modeling the 
transition amplitudes for intermediate hadronic states.

\section{Coherent vector meson production} 
\label{sec:coherent}

In the remainder of the paper we focus on the coherent 
production of vector mesons, i.e. the deuteron target 
stays intact. 
Note that our results can be easily generalized to 
vector meson production with deuteron break-up, as well as 
to exclusive pion production processes, 
like $\gamma^* +d \rightarrow \pi^-+ p p$, 
where color coherence  effects can also be explored \cite{CFS}.

Before we discuss specific reactions  suited to investigate 
re-scattering processes, 
we remind on basic features of the leptoproduction cross 
section and  vector meson dominance.

\subsection{Differential cross section and vector meson dominance} 

The coherent vector meson production amplitude in 
(virtual) photon-deuteron interactions can be obtained from 
Eqs.(\ref{eq:amplitude}) and (\ref{eq:ampl_ii}) 
if one requires the final $pn$ state to be a deuteron 
($f=d$).
If the polarization of the vector meson is not observed 
the leptoproduction cross section reads:
\begin{equation}\label{eq:cross}
\frac{d\sigma^{s,mm'}_{l d \rightarrow l V d}}{d Q^2 d\nu d t d\phi} = 
{\Gamma_V\over 32 \pi^2} L^{\mu\nu}
\left(F^{\rho}_{\mu}F^{\dag}_{\nu \rho}\right)^{s,mm'}.
\label{dif_crs}
\end{equation}
Here we have specified the spin quantum numbers which were 
labeled in the previous section by $j$ and $j'$. 
The index $s$ specifies the deuteron spin quantization axis and  
$m,\,m' = 0,\pm 1$ are the corresponding spin projections 
of the target before and after the scattering, respectively.
The four-momenta of the incoming and scattered lepton are 
$k_e^{\mu} = (E_e, {\vek k_e})$ and  $k_e^{\mu '} = (E_e',{\vek k_e'})$.   
$L^{\mu\nu}= {1\over 2}Tr(\hat k_e'\gamma^\mu \hat k_e \gamma^\nu)$
stands for the leptonic tensor with $\hat k = k_{\mu} \gamma^{\mu}$,  
neglecting terms proportional to the lepton mass.

Furthermore, $\phi$ denotes the angle between the lepton scattering plane 
defined by the momenta $\vek k_e$ and $\vek k_e'$, and the 
vector meson production plane  determined by the 
photon and vector meson momenta  $\vek q$ and $\vek k_V$.
Finally, $\Gamma_V$ is related to the flux of the virtual photon:
\begin{equation}
\Gamma_V = {\alpha_{em}\over 2\pi}{K\over Q^2}{1\over E_e^2}{1\over
  1-\epsilon},
\label{flux}
\end{equation}
where $\alpha_{em}=1/137$ is the electromagnetic coupling 
constant, $K=\nu(1-x)$, and $\epsilon= (4 E_e E_e' - Q^2)/(2 (E_e^2 + E_e'^2) 
+ Q^2)$ specifies the photon polarization.
Multiplying the tensor $F_{\mu\rho}$ in (\ref{dif_crs}) 
with the polarization vectors of the incident photon   
and the  produced vector meson yields the  photoproduction 
amplitude 
$F_d = \epsilon_{\gamma^*}^{\mu} F_{\mu\nu} \epsilon_{V}^{\nu}$ from 
Eqs.(\ref{eq:amplitude}) and (\ref{eq:ampl_ii}).

The differential cross section  (\ref{dif_crs}) 
is often written in terms of structure functions with definite 
helicity \cite{FSC}. 
For this purpose the following photon polarization vectors 
are introduced: 
\begin{eqnarray}
\epsilon_{\pm}^{\mu} & = & \mp{1\over\sqrt{2}} 
\left(0,1,\pm i,0\right),
\\
\epsilon_{0}^{\mu} & = & {1\over \sqrt{Q^2}}
\left(\sqrt{\nu^2 + Q^2},{\bf 0}_\perp,\nu\right).
\label{gpol}
\end{eqnarray}
Note that in the above definitions  the $z$-axis has been chosen parallel 
to the photon momentum. 
The differential cross section can then be written as: 
\begin{equation}
\frac{d\sigma^{s,mm'}_{l d \rightarrow l V d}}{d Q^2 d\nu dt d\phi}  =  
{\Gamma_V\over 32 \pi^2}
\left( \sigma_T^{s,mm'} + \epsilon\,\sigma_L^{s,mm'} - 
\epsilon \,\cos(2\phi ) \,\sigma_{TT}^{s,mm'} + \sqrt{\epsilon (1+\epsilon)} 
\cos(\phi)\,\sigma_{TL}^{s,mm'}\right). 
\label{dif_crs_hlc}
\end{equation}
Here the different structure functions are defined as: 
\begin{eqnarray}
\sigma_T^{s,mm'} & = &  {1\over 2}
\left(F^{\rho}_{+}F^{\dag}_{+\,\rho} + 
F^{\rho}_{-}F^{\dag}_{-\,\rho} \right)^{s,mm'},  
\nonumber \\
\sigma_L^{s,mm'} & = &  
\left(F^{\rho}_{0}F^{\dag}_{0\,\rho}\right)^{s,mm'}, 
\nonumber \\
\sigma_{TT}^{s,mm'} & = & \left(Re \,
F^{\rho}_{+}F^{\dag}_{-\,\rho}\right)^{s,mm'}, 
\nonumber \\
\sigma_{TL}^{s,mm'} & = & \left( Re \,F^{\rho}_{0} 
\left(F_{+ \,\rho} - F_{-\,\rho}\right)^{\dag}\right)^{s,mm'},
\label{str_fn}
\end{eqnarray}
where $F_{\lambda \rho}^{s,mm'} = 
\epsilon_{\lambda}^{\mu} F_{\mu \rho}^{s,mm'}$, 
with the helicity $\lambda = 0,+,-$.

The leptoproduction cross section (\ref{dif_crs_hlc}) 
is connected to the virtual photoproduction cross section 
via:          
\begin{equation}
\frac{d\sigma^{s,mm'}_{\gamma^* d \rightarrow V d}}{ dt d\phi} = 
\frac{1}{\Gamma_V} 
\frac{d\sigma^{s,mm'}_{l d \rightarrow l V d}}{d Q^2 d\nu dt d\phi}. 
\label{dif_crs_pho1}
\end{equation}
If the cross section is integrated over the azimuthal angle $\phi$   
the helicity flip structures $\sigma_{TT}$ and $\sigma_{TL}$ drop 
out and we end up with:
\begin{equation}
\frac{d\sigma^{s,mm'}_{\gamma^* d \rightarrow V d}}{dt} = 
{1 \over 16 \pi} \left( \sigma^{s,mm'}_T + \epsilon\sigma^{s,mm'}_L\right).
\label{dif_crs_pho2}
\end{equation}
It  is an experimental fact that in vector 
meson production from free nucleons 
at small $|t|$ the helicity of the vector meson 
is, to a good approximation, equal to the helicity of the incoming 
photon \cite{SCHC}.
Assuming this so-called $s$-channel helicity conservation for both 
amplitudes $f^{\gamma^*N\rightarrow hN}$ and $f^{hN\rightarrow VN}$ 
in (\ref{eq:amplitude}) yields\footnote{The 
quality of $s$-channel helicity conservation can be investigated according to 
(\ref{dif_crs_hlc}) by out-of-plane measurements.} 
$\sigma_{TT}=\sigma_{TL}=0$.
The helicity conserving structures $\sigma_T$ 
and $\sigma_L$ in (\ref{str_fn}) can be calculated from the 
photon-deuteron amplitude in Eq.(\ref{eq:amplitude})
if the nucleon amplitudes $f^{\gamma^* N\rightarrow hN}$ 
and $f^{h N\rightarrow VN}$ are specified.
Here this is done within the framework of 
the vector meson dominance model, 
where the electromagnetic current which enters in  
(\ref{eq:ampl_ii}) is represented by vector meson  fields 
(see e.g. \cite{BSYP} and references therein):
\begin{equation}
J_\mu^{em} = \sum_V {\sqrt{\pi \alpha_{em}} m^2_V\over g_V}
\delta(m^2_{h'}-m^2_V)V_\mu, 
\label{j_vmd}
\end{equation}
where $V_{\mu}$ ($V=\rho,\omega,\phi$) stands for the 
vector meson field with invariant mass $m_V$ and coupling constant 
$g_V$.
Combining Eqs.(\ref{eq:spectral}),(\ref{eq:en_den}) and (\ref{j_vmd})
yields the vector meson production amplitude for transversely 
polarized photons:
\begin{equation}
f^{\gamma^*_T N \rightarrow V N} =  
\sum\limits_{V'}{\sqrt{\alpha_{em}\pi}\over g_{V'}}
{m^2_{V'}\over m^2_{V'}+Q^2}
f^{V'_T N\rightarrow  VN} \approx 
{\sqrt{\alpha_{em}\pi}\over g_{V}}{m^2_{V}\over m^2_{V}+Q^2}
f^{V_T N\rightarrow  V N}.
\label{eq:VMDT}
\end{equation} 
In the following we neglect off-diagonal transitions 
$V'N\rightarrow V N$.  
In analogy with the suppression of inelastic diffraction to $\pi'$ 
states in diffractive pion-nucleon scattering, 
$\pi N \rightarrow X N$, they are expected to be
small.
Furthermore, since the  deuteron is an isosinglet 
the diagonal approximation is justified  in the kinematic region 
where vector meson dominance is applicable at least for coherent 
$\rho$-production due to the  different isospin of the 
$\rho$- compared to the $\omega$- and $\phi$-meson. 
Consequently, the diagonal approximation is exact for 
single scattering contributions while off-diagonal contributions 
are expected to be small for double scattering.\footnote{
Note that at large $Q^2$ non-diagonal transitions to states with 
invariant masses $m_h^2 \lsim  Q^2$ are important and 
are supposed to lead to color transparency.}

In vector meson dominance the production amplitude for longitudinally 
polarized virtual photons is related to the corresponding 
amplitude for transverse photons by (see e.g.\cite{FSC}):
\begin{equation}
f^{\gamma^*_L N\rightarrow  V N} = \xi {\sqrt{Q^2}\over m_V}
 f^{\gamma^*_T N\rightarrow  V N}, 
\label{eq:VMDL}
\end{equation}
where the ratio of the longitudinal to transverse vector meson-nucleon 
amplitude is denoted by
$\xi =  f^{V_L N\rightarrow  VN}/f^{V_T N\rightarrow  V N}$.

It remains to fix the amplitude $f^{\gamma^*_T N \rightarrow V N}$.
We concentrate on diffractive $\rho$-production and use 
for explicit calculations the parameterization: 
\begin{equation} \label{eq:prod_ampl}
f^{\gamma^*_T N \rightarrow V N} = \sigma_{\gamma^*\rho} 
(i +  \alpha) e^{{B(Q^2)\over 2} t}, 
\end{equation}
where $t\approx -\vek k^2_t$.
In real photoproduction at $\nu = 17\,GeV$ one finds 
typically $\sigma_{\gamma \rho}\approx  68\,\mu b$ \cite{BSYP}. 
For the slope $B(Q^2)$ we use the empirical values 
\cite{BSYP,slopes}:
$B(Q^2 < 1\,GeV^2) = 7\,GeV^{-2}$, 
$B(1\,GeV^2 < Q^2 < 2\,GeV^2) = 6\,GeV^{-2}$  and  
$B(2\,GeV^2 < Q^2 < 10\,GeV^2) = 5\,GeV^{-2}$. 
The ratio of real to imaginary part of the production amplitude 
in (\ref{eq:prod_ampl}) is fixed at $\alpha \approx -0.2$,  
and  we use $\xi^2 = 0.5$ \cite{BSYP}. 
The $\rho N$-amplitude is then obtained from Eq.(\ref{eq:VMDT}) 
using the  slope $B\approx 8~GeV^{-2}$. 
Note that at  intermediate energies the real part of the scattering 
amplitudes for proton and neutron targets are not exactly the same. 
We omit this difference since coherent vector meson production from 
deuterium is dominated by isospin averaged amplitudes.

In the following we consider the photoproduction cross section 
(\ref{dif_crs_pho2}).
It is determined by the square of the production amplitude 
Eq.(\ref{eq:amplitude}). In the framework of vector 
meson dominance the latter is given explicitly in  
Appendix \ref{appendix_B}.

\subsection{Evidence for double scattering}  
\label{3a}

Our main goal is to find kinematic domains  where 
the vector meson production cross section is sensitive 
to the re-scattering amplitude (Fig.1b).
First however let us recall that there is 
experimental evidence for double scattering which is 
theoretically well understood. 
In Fig.2 we show data on coherent photoproduction of 
$\rho$-mesons from unpolarized deuterium measured at 
SLAC \cite{Overman,Anderson} for a photon energy $\nu = 12\,GeV$. 
We compare the experimental data with results obtained from  
Eqs.(\ref{dif_crs_pho2}) and (\ref{eq:VMDT}).

It is evident that vector meson dominance successfully describes 
the measured data. 
Furthermore, one observes for  $ -t \gsim 0.4\,GeV^2$ significant 
contributions from double scattering.
Here the single scattering (Born) contribution 
as well as the contribution from the interference 
of single and double scattering become less important. 
At $ - t>0.6\,GeV^2$ the differential cross section 
is entirely controlled by double scattering. 

The importance of double scattering at large transfered momenta 
can easily be understood investigating 
the corresponding amplitudes in (\ref{eq:ia}) and (\ref{eq:DS}).
At large $|t|$ single scattering is suppressed through 
the large momentum transfer which enters in the deuteron form factor. 
In double scattering, however, the transfered 
momentum is shared between both interacting nucleons. 
Therefore  re-scattering probes 
the deuteron form factor at moderate momenta even at large $|t|$.
In the present analysis we restrict ourselves to $|t| < M_d^2$  
where corrections due to relativistic components in the 
deuteron wave function are expected to be small.

In the following we discuss further options to 
probe  double scattering contributions.
Especially polarized deuterium targets 
provide various  possibilities for their investigation as we will 
demonstrate.
Before focusing on signatures for color coherence we 
investigate  effects sensitive to the characteristic longitudinal 
interaction length of the production process.

\subsection{Finite longitudinal interaction length} 
\label{3b}

As already mentioned in the introduction high-energy 
(virtual) photon-induced processes are dominated  by contributions from 
large longitudinal space-time intervals which increase with 
the energy of the incident photon. 
For coherent vector meson production the characteristic 
distances can be extracted from the production  
amplitude in Eq.(\ref{eq:amplitude})
after a Fourier transformation into coordinate space.  
The obtained length scales are to a good approximation  
inversely proportional 
to the longitudinal momentum transfers which enter in 
the form factors in Eq.(\ref{eq:amplitude}). 
For single scattering one finds: 
\begin{equation} \label{eq:long_Born}
\delta_z^{(a)} \sim \left|\frac{1}{l_{-}}\right| = 
\frac{2\nu}{Q^2 + m_V^2 - t}, 
\label{l1}
\end{equation}
while for double scattering one gets:
\begin{equation} \label{eq:long_ds}
\delta_z^{(b)} \sim \left|\frac{1}{-2\Delta_V}\right| 
= \left|\frac{2\nu}{Q^2 + m_V^2 + t}\right|. 
\label{l2}
\end{equation}
Note that if the deuteron recoil is neglected  
the longitudinal interaction length for  single scattering 
reads: 
\begin{equation}
\delta_z^{(a)}\sim\left| {1\over l_z}\right| = 
{2\nu \over Q^2 + m_V^2 - t -\nu t/M_d} 
\stackrel{\nu \rightarrow \infty }{\llongrightarrow}
2M_d/\left|t\right|, 
\label{l0}
\end{equation}
which for $t\ne 0$ becomes a constant at high energies.
This contradicts our basic understanding of high-energy 
photon-induced processes being controlled by longitudinal 
distances which increase with the photon energy $\nu$ 
\cite{GIP,Gribov70,LL,Ioffe}. 

At intermediate energies where characteristic longitudinal 
interaction distances are of the order of the target size, 
$\delta_z^{(a)}, \delta_z^{(b)} \sim \langle r^2\rangle^{1/2}_d$, 
a strong energy dependence of the cross section is expected. 
In momentum space this can be traced back to the sensitive 
momentum dependence of the deuteron form factor 
which enters  the production amplitude (\ref{eq:amplitude}).

For a quantitative investigation we consider 
the $\rho$-meson photoproduction cross section  (\ref{dif_crs_pho2})
for an unpolarized deuterium target 
using proper longitudinal interaction 
distances (\ref{l1}, \ref{l2}) as they appear in the scattering
amplitude (\ref{eq:amplitude}), normalized by the cross section 
calculated in the limit $\delta_z^{(a)}, \delta_z^{(b)} \rightarrow \infty$:
\begin{equation}
R_{\delta_z} = 
\left.
\frac{d\sigma_{\gamma^* d\rightarrow \rho d}}{dt}
\right/
\frac{d\sigma_{\gamma^*d\rightarrow \rho d}}{dt}(l_{-} = \Delta_{\rho} = 0).  
\label{Rl}
\end{equation}
In Fig.3 we show $R_{\delta_z}$ for photo- and 
leptoproduction in the kinematic domain of  TJNAF  \cite{LOI}. 
Here the characteristic longitudinal interaction distance   
for single scattering is typically of the order of the deuteron size. 
For example at $\nu = 4\,GeV$ and $t = -0.2\,GeV^2$,
one finds $\delta_z^{(a)} \approx 2\,fm$. 
Consequently we observe at small $-t\lsim 0.4\,GeV^2$, where the Born 
term dominates, a strong dependence of the production cross section on 
$\nu$ and $Q^2$.
Indeed at $t \sim -0.1 \,GeV^2$
an increase in the photon energy\footnote{Note 
that the validity of vector meson dominance \cite{BSYP} demands 
a minimal photon energy $\nu \gsim 3\,GeV$.} from $3\,GeV$ to $6\,GeV$
leads to a $15\%$ rise of $R_{\delta_z}$.
In leptoproduction at $ t \sim - 0.1\,GeV^2$ and $\nu=6 \,GeV$ 
an increase of $Q^2$ from $0.5\,GeV^2$ to $2\,GeV^2$ yields 
a decrease of $R_{\delta_z}$ by approximately $50\%$.
At large $-t> 0.7\,GeV^2$, where double scattering dominates, only minor 
variations of the production cross section occur 
since here smaller proton-neutron distances are probed.

If vector meson production is considered at higher photon energies  
effects from finite longitudinal  interaction distances occur too, 
however at larger values of $Q^2$. 
As an example we present in Fig.4 the 
cross section ratio $R_{\delta_z}$ for 
$\nu = 30 \,GeV$ and $Q^2 = (2-10)\,GeV^2$.
This kinematic region is accessible at HERMES \cite{HC}.
Although at $Q^2>2\,GeV^2$ vector meson dominance does not describe 
the $Q^2$-dependence of the nucleon production amplitude 
$f^{\gamma^*N\rightarrow VN}$ properly, in the ratio $R_{\delta_z}$
to a good approximation any $Q^2$-dependence of 
the single scattering amplitude drops out.
Therefore in the domain where the Born contribution dominates, i.e. 
at $-t\lsim 0.4\,GeV^2$, the presented results should be reasonable.  
As before we observe at  $ t\sim - 0.1\,GeV^2$ 
a strong rise of the production cross section for decreasing 
values of $Q^2$. 
Double scattering,  which dominates the cross section at 
large $|t|$, will depend sensitively on $Q^2$ due to color 
coherence effects.
Therefore  at large $Q^2>2\,GeV^2$ and $-t>0.4\,GeV^2$ the cross 
section ratio  $R_{\delta_z}$ calculated within vector meson dominance 
should be considered  as a baseline estimate assuming 
color coherence effects to be absent.

Coherent vector meson production from polarized deuterons 
provides further possibilities to investigate effects due 
to a change of the characteristic longitudinal interaction 
length.
If the polarization of the scattered deuteron is not 
observed  the Born cross section is proportional 
to the deuteron density matrix (see Appendix B and C): 
\begin{equation} \label{eq:density_matrix}
\rho^{s,m}\left( \frac{\vek l_{\perp}}{2},-\frac{l_{-}}{2}\right)  
= 
\sum_{m'} 
S_d^{s, m m'}\left(\frac{\vek l_{\perp}}{2},-\frac{l_{-}}{2}\right)
S_d^{s,m m'}\left(\frac{\vek l_{\perp}}{2},-\frac{l_{-}}{2}\right)^{\dag}. 
\end{equation}
Note that $l_{\_}$ enters on the left- and right-hand side of 
Eq.(\ref{eq:density_matrix}) as a consequence of the 
recoil of the final state (Appendix \ref{appendix_A}).
Choosing the spin quantization axis parallel to the photon momentum 
$\vek q$, which we denote by $s=1$, one finds for  
$m=0$ (\ref{eq:density_matrix_explicit}): 
\begin{equation}
\rho^{1,0} \left(\frac{\vek l_{\perp}}{2},-\frac{l_{-}}{2}\right)
= F_C^2(\tilde l/2) +  
\left[{3l^2_{-}\over \tilde l^2}-1\right]
\sqrt{2}F_C(\tilde l/2)F_Q(\tilde l/2) + 
\left[{3l^2_{-}\over \tilde l^2}+ 1\right]{F_Q^2(\tilde l/2)\over 2},
\end{equation}
where $\tilde l^2 = \vek l_{\perp}^2 + l_{-}^2$. 
$F_C$ and $F_Q$ are the deuteron monopole and quadrupole 
form factors as defined in the appendix (\ref{eq:form_factors}).
For the Paris potential \cite{LaLoRi80} they are shown in Fig.5.  
In the limit $l_{-}=0$ the density matrix reduces to:
\begin{equation} \label{eq:F_C - F_Q}
\rho^{1,0} \left(\frac{l_{\perp}}{2},0\right)
= \left(F_C(l_{\perp}/2) - \frac{1}{\sqrt{2}} F_Q(l_{\perp}/2)\right)^2.  
\end{equation}
with $l_{\perp} = |\vek l_{\perp}|$. From 
Fig.6 we observe that $\rho^{1,0}(l_{\perp}/2,0) =  0$ 
for $l_{\perp} \approx 0.7\,GeV$,   
leading to the disappearance of the single scattering 
contribution.
Consequently  for  $l_{-}\ll \sqrt{-t/3}$ 
the production cross section is dominated  by double scattering 
contributions at $t \approx -l_\perp^2 \approx - 0.5\,GeV^2$.
We therefore expect in this region of $t$ a strong energy dependence of 
the Born contribution since $l_{-}\sim 1/\delta_z^{(a)} \sim 1/\nu$.  

In Fig.7  we show the $\rho$-meson photoproduction cross section 
(\ref{dif_crs_pho2}) 
\begin{equation} 
\frac{d\sigma_{\gamma^* d\rightarrow \rho d}^{1, 0}}{dt} =
\sum_{m'}\frac{d\sigma_{\gamma^* d\rightarrow \rho d}^{1, 0 \,m'}}{dt}.
\end{equation}
for different photon energies. 
We indeed observe a  significant energy dependence of the Born 
contribution at $t\approx  - 0.5 \,GeV^2$. 
Combined with  double scattering it leads  
to a decrease of the production cross section at 
$t \approx  -0.35\,GeV^2$ of more than two orders of magnitude 
if the  photon energy rises from $\nu = 3\,GeV$ to $10\,GeV$.

\subsection{Color coherence effects at large $|t|$} 
\label{3c}
 
In Sec.\ref{3a} we have demonstrated that within the 
framework of vector meson dominance double scattering dominates 
the coherent production cross section at $-t > 0.6\,GeV^2$. 
However, as outlined in the introduction, at  large $Q^2 \gg 1\,GeV^2$ 
the hadronic state which is formed  in the interaction of the 
virtual photon with a nucleon is in general not a vector meson 
as in vector meson dominance, but a quark-gluon wave 
packet with a characteristic transverse size $b_{ej} \sim 1/Q$. 
If the energy of this  ejectile is large enough,  such that its 
formation time $\tau_f \approx 2\nu/\delta m_V^2$ exceeds the  
target size, it will be the ejectile which re-scatters from 
the second nucleon. 
One therefore expects double scattering to vanish at 
$Q^2 \gg 1\,GeV^2$ and $\nu > \langle r^2 \rangle_d^{1/2} \delta m_V^2/2 
\approx 10\,GeV$. 
This  color coherence or color transparency effect 
can be investigated through the $Q^2$-dependence of the 
vector meson production cross section in kinematic regions 
where it is most sensitive to re-scattering.
However, for this purpose it is mandatory to account for 
eventual modifications 
due to a change of the characteristic longitudinal interaction 
lengths  $\delta_z^{(a)}$ and  $\delta_z^{(b)}$ from 
Eqs.(\ref{l1},\ref{l2}). 
Or, even better, the kinematics should be chosen such 
that these length scales stay constant.

In Fig.8 we present a baseline calculation for coherent 
$\rho$-production from unpolarized deuterons 
within vector meson dominance. 
For different values of $Q^2$ 
the corresponding photon energies $\nu$ 
are chosen such that $x = 0.1$. 
At large values of $\nu$ we then have  
$\delta_z^{(a)} \approx \delta^{(b)} \approx 1/Mx \approx 2 \,fm$.
We normalize the production cross section by the cross section 
taken at 
$t=t_{min}$ and show the ratio:
\begin{equation}
R = 
\left.\frac{d\sigma_{\gamma^*d \rightarrow \rho d}}{dt}
\right/
\frac{d\sigma_{\gamma^*d \rightarrow \rho d}}{dt} (t=t_{min}).
\label{R_Q2}
\end{equation}
Note that in $R$ 
any $Q^2$- and $\nu$-dependence of the initial photoproduction 
process cancels.
In addition we show in Fig.8 the corresponding cross section ratio 
$R_{Born}$ which accounts for single scattering only. 
We observe that at large $-t \gsim (0.8-1)\,GeV^2$ the full production 
cross section is approximately a factor $2.5$--$5$ larger than the 
Born contribution. 
This difference is the maximal effect which can be caused 
by color coherence: at small $Q^2 < 1\,GeV^2$ 
vector meson dominance is applicable and the hadronic 
state which re-scatters can be represented by a soft vector mesons. 
Consequently the cross section ratio should be identical to $R$. 
However if, as a consequence of color coherence, double scattering 
is absent at large $Q^2$ and large $\nu$, 
only single scattering contributes to the production cross section and 
yields the ratio $R_{Born}$.

The difference between the Born cross section and the 
full production cross section in absence of color coherence 
can be enhanced, if one considers the ratio of the  
cross sections at large and moderate $|t|$. 
In absence of color coherence vector meson production at large 
$|t|$ is dominated by double scattering and the overall cross 
section is larger than the Born contribution. 
However at moderate $t\sim - 0.4\,GeV^2$ the interference of the  
double and single scattering amplitude is important   
and leads to an overall production cross section being 
smaller than the Born term.
Taking the ratio of both gives an enhanced sensitivity to the  
double scattering contribution and thus to eventual 
color coherence effects.
Furthermore in such a cross section ratio any $Q^2$-dependence, 
apart from being caused by color coherence, cancels to a large 
extend.  
In Fig.9 we present the ratio of the $\rho$-meson production 
cross sections from unpolarized deuterons at $x=0.1$ for 
$t = -0.4\,GeV^2$ and $t=-0.8\,GeV^2$. 
The Born cross section and the full vector meson dominance 
calculation differ by a factor of $4$,  
leaving reasonable room for an investigation of color coherence.

\subsection{Color coherence at moderate $|t|$} 
\label{3d}

At moderate values of $|t|$ vector meson production from unpolarized 
deuterons is largely dominated by the Born contribution.
For polarized targets, however, even here  
kinematic windows exist where the Born contribution is small and 
the production cross section becomes sensitive to re-scattering.

We first consider a deuteron target with spin projection $m=0$ 
along the photon momentum $\vek q$ (labeled as previously by $s=1$). 
For such a polarization we have found in  Sec.\ref{3b} 
a  dip in the Born cross section  at $t \approx -0.5\,GeV^2$.
The latter, however, depends strongly on $l_{-}$ or, 
equivalently, the longitudinal interaction length $\delta_z^{(a)}$.
To suppress this dependence we consider the ratio:  
\begin{equation} \label{eq:R10}
R^{1,0}  = 
\left.\frac{d\sigma_{\gamma^* d\rightarrow \rho d}^{1, 0}}{dt}
\right/
\frac{d\sigma_{\gamma^* d\rightarrow \rho d}^{1,0}}{dt} 
\left(t=t_{min}\right),
\end{equation}
for various $Q^2$ but at fixed values of $x$.
In Fig.10 we compare $R^{1,0}$ with the corresponding cross section ratio 
$R_{Born}^{1,0}$ obtained from single scattering only.
As expected, at small $x$  the Born contribution develops a minimum 
at $t \approx - 0.5 \,GeV^2$.
Here we find at $x=0.05$  an order of magnitude difference between 
the Born and the full vector meson dominance cross section.
At $x=0.001$ the difference is even larger than two orders of 
magnitude. 
Therefore an investigation of the $Q^2$-dependence of the 
cross section ratio $R^{1,0}$ in the region $t\approx -0.5\,GeV^2$ 
can serve as a tool to investigate double 
scattering contributions to vector meson production and thus 
color coherence.

An even more interesting choice for the target polarization is to 
take the spin projection $m=0$ perpendicular to the 
vector meson production plane, 
i.e. parallel to $\vek \kappa = \vek q\times \vek l$, 
which we label as $s=2$.  
The corresponding deuteron density matrix
for the Born cross section reads (\ref{den}):
\begin{equation}
\rho^{2,0}\left(\frac{\vek l_\perp}{2},-\frac{l_{-}}{2}\right) 
= \left(F_C^2(\tilde l/2 ) - \frac{1}{\sqrt{2}} F_Q(\tilde l/2) 
\right) ^2,
\label{deny}
\end{equation}
irrespective of the magnitude of the longitudinal momentum 
transfer $l_{-}$. 
Consequently, for this target polarization a node in the 
Born contribution occurs for $\tilde l = 0.7\,GeV$,  
which again corresponds 
for the large energies considered approximately 
to  $t \approx - 0.5\,GeV^2$.
In Fig.11 we show the $t$-dependence of the cross 
section ratio $R^{2,0}$ defined similar to   Eq.(\ref{eq:R10}). 
The observed node in the Born contribution leads to a 
perfect window for an  investigation of  double scattering.

Finally we consider the cross section for tensor polarization 
normalized by the unpolarized production cross section:
\begin{equation}\label{eq:tensor_pol}
A_d = \frac{d\sigma^{2,1}/dt + d\sigma^{2,-1}/dt - 
      2 d\sigma^{2,0}/dt}
      {d\sigma/dt},
\end{equation}
where the spin quantization axis is taken parallel to  $\vek \kappa$. 
In comparison to the corresponding asymmetry for the Born contribution 
we find a significant sensitivity to double scattering 
at $-t > 0.6\,GeV^2$ as demonstrated in Fig.12.

\section{Conclusions} 
\label{sec:conclusions}

Coherent leptoproduction of vector mesons from deuterium provides
unique possibilities to investigate the characteristic longitudinal 
interaction length in high-energy 
photon-induced processes, and 
to search for color coherence phenomena.
We have presented a derivation of the corresponding amplitudes 
including the recoil of the scattered deuteron. 
The latter is neglected in  common derivations based on 
the Glauber multiple scattering formalism, but 
turns out to be important in the case of 
non-forward production processes from light nuclei.

We have identified kinematic regions where  vector meson 
production is influenced significantly by   
variations of the longitudinal interaction length of the photon 
which  depends in a specific way  on the 
photon energy and the momentum transfer. 
Several possibilities  for an investigation of the latter in 
the kinematic domain of TJNAF and HERMES are outlined.

A further issue of this work was the identification of 
kinematic windows where the vector meson production cross 
section is most sensitive to re-scattering contributions.
Here the propagation of small size quark-gluon configurations, 
which are initially produced in high-energy lepton-nucleon 
interactions at large $Q^2$, can be studied.
We found promising signatures for color coherence in the 
kinematic domain of HERMES. 
For unpolarized deuteron targets they involve large momentum 
transfers $-t\gsim 0.6\,GeV^2$, 
while for polarized targets re-scattering can be investigated 
already at moderate $-t \sim 0.5 \,GeV^2$.

\begin{acknowledgements}

This work was supported  in part by the German-Israeli 
Foundation Grant GIF -I-299.095, by the U.S. Department of Energy
under Contract No. DE-FG02-93ER40771, by the
National Science Foundation under Grants Nos. PHY-9511923 and PHY-9258270,
and by the BMBF.
M. Sargsian thanks the Alexander von Humboldt Foundation for support. 

\end{acknowledgements}

\newpage

\appendix

\section{Nuclear recoil}
\label{appendix_A}

In collisions at finite momentum transfer, $t = l^2 \ne 0$,
a proper consideration of the longitudinal interaction length 
(\ref{eq:long_Born}) and (\ref{eq:long_ds})  
goes beyond the conventional Glauber multiple scattering 
formalism which neglects nuclear recoil, i.e.
the energy difference of the nucleus before and after 
the interaction (see e.g. \cite{CheWu87}). 
As a consequence of the nuclear recoil  the arguments of the scattered 
two-nucleon wave function in the amplitudes (\ref{eq:ia}) and (\ref{eq:ds}) 
depend on $l_- = l_0-l_z$ instead of $l_z$ as in the conventional 
Glauber approximation.

To clarify this issue recall that high-energy processes 
develop near the light-cone $|z-t|\approx 0$.
Therefore they are naturally described in a light-cone description  
where  the ``$-$'' and the transverse momentum components are 
conserved. 
As a result the deuteron wave function depends on the  light-cone
fraction of the deuteron momentum carried by the interacting nucleon,
$\alpha = {p_{d-}-p_{s-}\over  p_{d-}}$,  and on $\vek p_{s\perp}$,      
where $p_d$ and $p_s^{\mu} = (p_{s0},\vek p_{s\perp},p_{sz})$
are the four-momenta  of the deuteron and the spectator nucleon,
respectively (Fig.1).

We consider the scattering process in a frame where the 
deuteron target is at rest.
In addition we take the momentum of the virtual photon along 
the $z$-axis. 
The invariant mass of the initial two-nucleon system is: 
\begin{equation} \label{llc1b}
M_i^2=\frac{M^2 + \vek p_{s\perp}^{2}}{\alpha(1-\alpha)}, 
\end{equation}
where $M$ denotes the nucleon mass. 
For the single scattering process  
the  mass of the two-nucleon system after the collision is: 
\begin{equation} 
M_f^2 = {M^2+ \vek p_{s\perp}^{2}\over 1- \alpha_f} + 
 {M^2+(\vek l_{\perp}-\vek p_{s\perp})^{2}\over \alpha_f} - 
\vek l_{\perp}^2 = 
{M^2 +  (\alpha_f \vek l_{\perp}- \vek p_{s\perp})^2
\over \alpha_f(1-\alpha_f)}. 
\label{llc2}
\end{equation}
where the final state, two-nucleon system is characterized by the variables:
\begin{equation}
\alpha_f = 1-{p_{s-} \over p_{d-} + l_{-}}, \hspace*{1cm} 
\mbox{and} \hspace*{1cm}  
\vek p_{f\perp} = \alpha_f \vek l_{\perp}-\vek p_{s\perp}.
\label{llc1}
\end{equation}
On the other hand  in the center of mass the 
invariant mass of the two-nucleon system before and after the interaction 
depends on the nucleon three-momentum only:  
$M^2_i\approx 4(M^2+\vek p_s^{2})$ and $M_f^2 \approx 4(m^2+ \vek p_f^2)$.
One then obtains in the non-relativistic limit:  
\begin{equation}
\alpha = {1\over 2}\left(1+{p_{sz}\over M}\right), \ \ \ \ 
\alpha_f = {1\over 2}\left(1 - {p_{fz}\over M}\right), \ \ \mbox{and} \ \ 
p_{fz} = -p_{sz} - {l_-\over 2}
\label{llc3}.
\end{equation}
Consequently the wave function of the target deuteron depends on 
$p_{sz}$ and  $\vek p_{s\perp}$,  while in the 
wave function of the final two-nucleon system  
$p_{fz} = - p_{sz} - {l_-}/2$ and  
$\vek p_{f\perp}  = - \vek p_{s\perp} + {1\over 2}(1 - {p_{fz}\over M})
\vek l_{\perp}\approx - \vek p_{s\perp} + \vek l_{\perp}/2$ 
enters. 
Note that $l_0 = \sqrt{M_{f}^2+\vek l^2}- M_i$ 
corresponds to the recoil energy of the target.

Similar considerations can be repeated for the re-scattering process 
shown in Fig.1b. 
Here the light-cone momentum fraction of the 
nucleon in the intermediate state (after interacting 
with the photon) is:
\begin{equation}
\alpha_f = {p_{d-}-p_{s-}+l_{-}-k_{-}\over p_{d-} + l_{-}} = 
{p_{d-}-p_{s-}+l_{-}-k_{-}\over p_{d-}}{p_{d-}\over  p_{d-} + l_{-}}.
\label{llc4}
\end{equation}
where $k^\mu=(k_0,{\vek k}) = p_{s}^{'\mu}-p_{s}^{\mu}$.
In the non-relativistic limit one obtains:
\begin{equation}
\alpha_f \approx  {1\over 2}\left(1-{p_{fz}\over M}\right) = 
{1\over 2}\left(1 + {p_{sz}\over M} + {l_{-}\over 2M} - {k_{-}\over M}\right).
\label{llc5}
\end{equation} 
We then get:
\begin{equation}
p_{fz}= -{l_{-}\over 2}+k_{-} - p_{sz},
\label{llc6}
\end{equation}
which corresponds to the longitudinal momentum component which enters 
the wave function in the double scattering amplitude in 
Eq.(\ref{eq:ds}).

Note that in a  light-cone approach which is necessary for 
a proper treatment of the production process at large momentum 
transfers $-t \ge (1-2)~GeV^2$  one needs to introduce 
relativistic deuteron wave functions.
The corresponding light-cone  transition form factor \cite{FS81}
naturally accounts for  the recoil of the final nuclear system:
\begin{equation}
S(\vek k_{\perp}, l_{-})= \int {d \alpha \over \alpha (1-\alpha)}
d^2{k_{\perp}} 
\psi_d(\alpha, \vek p_{\perp})
\psi_d(\alpha', \vek p_{\perp}-\vek k_{\perp} (1-\alpha')),
\end{equation}
where $\alpha'= (\alpha+l_{-}/M)/(1+{l_{-}\over 2 M})$.

\section{Photoproduction amplitudes}
\label{appendix_B}

Within the framework of vector meson dominance one obtains 
for the squared vector meson production amplitude 
for a deuterium target from 
Eqs.(\ref{eq:ia}) and (\ref{eq:DS}):
\begin{eqnarray}  \label{eq:squared_ampl}
\left|F^{jj'}_{d}\right |^2 = 
\left|F^{sm,m'}_{d}\right |^2  &=&  
F^{(a)}  F^{(a)\dag}  + 
2 Re\,\left(F^{(a)} \tilde F^{(b)\dag} \right) + 
\tilde F^{(b)}\tilde F^{(b)\dag} + \nonumber \\
& & 2 Re\,\left(F^{(a)}\Delta F^{(b)\dag} \right) + 
2 Re\,\left(\tilde F^{(b)}\Delta F^{(b)\dag} \right) + 
\Delta F^{(b)} \Delta F^{(b)\dag}.  
\end{eqnarray}
The different terms are: 
\begin{mathletters} \label{eq:ampl_coh}
\begin{eqnarray} 
&& F^{(a)}  F^{(a)\dag} = 4 \left| 
f^{\gamma^*N\rightarrow VN}(\vek l)\right| ^2 
S^{s,mm' }_{d}(\vek l_{\perp}/2,-l_{-}/2)^2, 
\nonumber \\ \\
&& 2Re ( F^{(a)}\tilde F^{(b)\dag}) =  
-2 Im \left\{ f^{\gamma^* N\rightarrow VN}(\vek l)
S_{d}^{s,mm'}({\vek l_{\perp}/2, -l_{-}/2})\right.
\nonumber \\
&&\hspace*{0.2cm}\times \left.
\int \frac{d^2 k_\perp}{(2 \pi)^2} 
f^{\gamma^*N\rightarrow VN} 
({ \vek l_{\perp}/2}-\vek k_\perp)^{\dag} 
f^{V N\rightarrow VN} 
({\vek l_\perp/2}+\vek k_\perp)^{\dag}
S_{d}^{s,mm'} (\vek k_\perp,-\Delta_V)\right\},\nonumber \\ 
\\
&& \tilde F^{(b)}\tilde F^{(b)\dag}   =   
\frac{1}{4} \int \frac {d^2 k_{\perp}}{(2 \pi)^2} 
\frac {d^2 k'_{\perp}}{(2 \pi)^2}  
\,f^{\gamma^*N\rightarrow VN}
({\vek l_{\perp}/ 2}-\vek k_{\perp})\,
f^{V N\rightarrow VN} 
({\vek l_{\perp}/2}+\vek k_{\perp})
\nonumber \\
&& \hspace{0.2cm}
\times
f^{\gamma^*N\rightarrow VN} 
({\vek l_{\perp}/2}-\vek k\,'_{\perp})^{\dag}\,
f^{V N\rightarrow VN} 
({\vek l_{\perp}/2}+\vek k\,'_{\perp})^{\dag}
S_{d}^{s,mm'} (\vek k_{\perp},-\Delta_V)  
S_{d}^{s,mm'} (\vek k'_{\perp},-\Delta_V),\nonumber \\ \\
&& 2 Re(F^{(a)}\Delta F^{(b)\dag})   =  
 -  Re 
\left\{
\sqrt{8 \over \pi}\,
f^{\gamma^* N\rightarrow VN}(\vek l)
S_{d}^{s,mm'}(\vek l_{\perp}/2,-l_{-}/2)
\right. 
\nonumber \\  
&& \left.\hspace{0.2cm} \times  
\int \frac{d^2 k_{\perp}}{(2 \pi)^2}
\,f^{\gamma^*N\rightarrow VN}
({\vek l_{\perp}/2}-\vek k_{\perp})^{\dag}\,
f^{V N\rightarrow VN} 
({\vek l_{\perp}/2}+\vek k_{\perp})^{\dag} 
\Delta S_{d}^{s,mm'} 
(\vek k_{\perp}, -\Delta_V)
\right\},\nonumber \\ \\
&&2Re(\tilde F^{(b)}\Delta F^{(b)\dag}) =  
-Im \,\left\{{1\over \sqrt{2\pi}}\,
\int \frac {d^2 k_{\perp}}{(2 \pi)^2} 
\frac {d^2 k'_{\perp}}{(2 \pi)^2} 
f^{\gamma^*N\rightarrow VN}
({\vek l_{\perp}/ 2}-\vek k_{\perp})\,
%f^{V N\rightarrow VN} 
%({\vek l_{\perp}/2}+\vek k_{\perp})
\right.
\nonumber \\
&& \hspace{0.2cm} \times \left.
f^{V N\rightarrow VN} 
({\vek l_{\perp}/2}+\vek k_{\perp})
f^{\gamma^*N\rightarrow VN}
({\vek l_{\perp}/2}-\vek k\,'_{\perp})^{\dag}\,
f^{V N\rightarrow VN} 
({\vek l_{\perp}/2}+\vek k\,'_{\perp})^{\dag} 
\right.
\nonumber \\
&& \hspace{0.2cm} \times \left.
S_{d}^{s,mm' }(\vek k_{\perp},  - \Delta_V)
\Delta S_{d}^{s,mm'}(\vek k\,'_{\perp},-\Delta_V)\right\},
\nonumber \\ \\
&& \Delta F^{(b)} \Delta F^{(b)\dag} = 
\frac{1}{2 \pi} \int 
\frac{d^2 k_{\perp}}{(2 \pi)^2}
\frac{d^2 k'_{\perp}}{(2 \pi)^2}
\,f^{\gamma^*N\rightarrow VN}
({\vek l_{\perp}/2}-\vek k_{\perp})\,
f^{V N\rightarrow VN} 
({\vek l_{\perp}2}+\vek k_{\perp})
\hfill \nonumber \hfill \\ 
&& \hspace{0.2cm} \times  
f^{\gamma^*N\rightarrow VN} 
({\vek l_{\perp}/2}-\vek k\,'_{\perp})^{\dag}\,
f^{V N\rightarrow VN} 
({\vek l_{\perp}/ 2}+\vek k\,'_{\perp})^{\dag}
\hfill \nonumber \hfill \\ 
&& \hspace{0.2cm} \times  
\Delta S_{d}^{s,mm'}(\vek k_{\perp},-\Delta_V)
\Delta S_{d}^{s,mm'}(\vek k\,'_{\perp},-\Delta_V). 
\end{eqnarray}
\end{mathletters}

\section{Deuterium form factor and density matrix}
\label{app:deuteron}

In coherent vector meson production processes 
the form factor which  enters in Eqs.(\ref{eq:ia},\ref{eq:DS}) 
is the elastic deuteron form factor. It can be split into 
a monopole and a quadrupole term   
(see e.g. Ref.\cite{BJ}): 
\begin{equation} 
S^{s,m m'}_d(\vek l) 
= \chi^{m \dag}_d \left(F_c(l) - F_Q(l)
\sqrt{{1\over 8}}\hat S(\vek l)\right)\chi^{m'}_d. 
\label{ff}
\end{equation}
Here $\chi^m_d$ is the deuteron spinor, 
$\hat S(\vek l) = {3(\vek \sigma_p\cdot  \vek l)
(\vek \sigma_n\cdot  \vek l)/ \vek l^2} - \vek \sigma_p \cdot \vek
\sigma_n$ 
the tensor operator 
and $l = |\vek l|$. 
The index $s$ specifies the spin quantization axis, while  
$m, m'  = 0,+,-$ label the spin projection of the 
deuteron before and after the interaction, respectively.  
The monopole and quadrupole form factor can be expressed in terms 
of $u$ and $w$, the radial $S$- and $D$-state wave functions of the 
deuteron:  
\begin{eqnarray} \label{eq:form_factors}
F_C(l) & = & \int\limits_0^\infty dr\,\left[u^2(r)+w^2(r)\right]j_0(lr),
\nonumber \\
F_Q(l) & = & \int\limits_0^\infty dr\,
\left[u(r)-{1\over \sqrt{8}}w(r)\right]
2\,w(r)j_2(lr).
\label{c10}
\end{eqnarray}
If the spin polarization of the final state deuteron is not observed 
the coherent scattering amplitudes can be written in terms of the 
density matrix \cite{AB}:
\begin{eqnarray} \label{eq:density_matrix_explicit}
\rho^{s, m}(\vek l_1,\vek l_2) &=& 
\sum_{m'} S_d^{s,m m'}(\vek l_1) S_d^{s,m m'}(\vek l_2)^{\dag} 
=  F_C(l_1) F_C(l_2)  
\nonumber \\
&+& 
\frac{1}{\sqrt{2}} 
\left\{ 
\left[
\frac{3|\vek l_2 \cdot \vek \epsilon_s^m|^2}{l_2^2} -1
\right] 
F_C(l_1) F_Q(l_2) + 
\left[
\frac{3|\vek l_1 \cdot \vek \epsilon_s^m|^2}{l_1^2} -1
\right] 
F_C(l_2) F_Q(l_1) 
\right. \nonumber \\
&+& 
\left.
\left[9\frac{(\vek l_1\cdot \vek \epsilon_s^m)
            (\vek l_1\cdot \vek \epsilon_s^m)^* 
             \vek l_1 \cdot \vek l_2}{l_1^2 \,l_2^2} 
- \frac{3|\vek l_1 \cdot \vek \epsilon_s^m|^2}{l_1^2} 
- \frac{3|\vek l_2 \cdot \vek \epsilon_s^m|^2}{l_2^2}  + 1
\right]
\frac{F_Q(l_1) F_Q(l_2)}{2}\right\}.
\label{den} 
\end{eqnarray}
Here $\vek\epsilon_s^m$ is the polarization vector of the target 
deuteron. 
For spin quantization parallel to the $z$ axis we have: 
\begin{equation} 
\vek\epsilon^{+} = 
\frac{-1}{\sqrt{2}}
\left(
\begin{array}{c} 1\\i\\0 
\end{array} 
\right), \quad
\vek\epsilon^{-} = 
\frac{1}{\sqrt{2}}
\left(
\begin{array}{c} 1\\-i\\0 
\end{array} 
\right),\quad
\vek\epsilon^0 = 
\left(
\begin{array}{c} 0\\0\\1 
\end{array} 
\right).
\label{mt}
\end{equation}             
Note that for practical purposes the representation of 
the spin density matrix in (\ref{eq:density_matrix_explicit}) 
in terms of the polarization 
vector $\vek\epsilon_s^m$ is very useful  since it is applicable 
for any choice of the spin quantization axis after rotating the 
polarization vectors appropriately.

%%%%%%%%%%%%%%%%%%%%%%%%%%%%%%%%%%%%%%%%%%%%%%%%%%%%%%%%%%%%%%%%%%

%%%%%%%%%%%%%%%%%%%%%%%%%%%%%%%%%%%%%%%%%%%%%%%%%%%%%%%%%%%%%%%%%%%%%%%%%

\newpage

\begin{figure}

\centerline{\epsfysize=8.0truein\epsffile{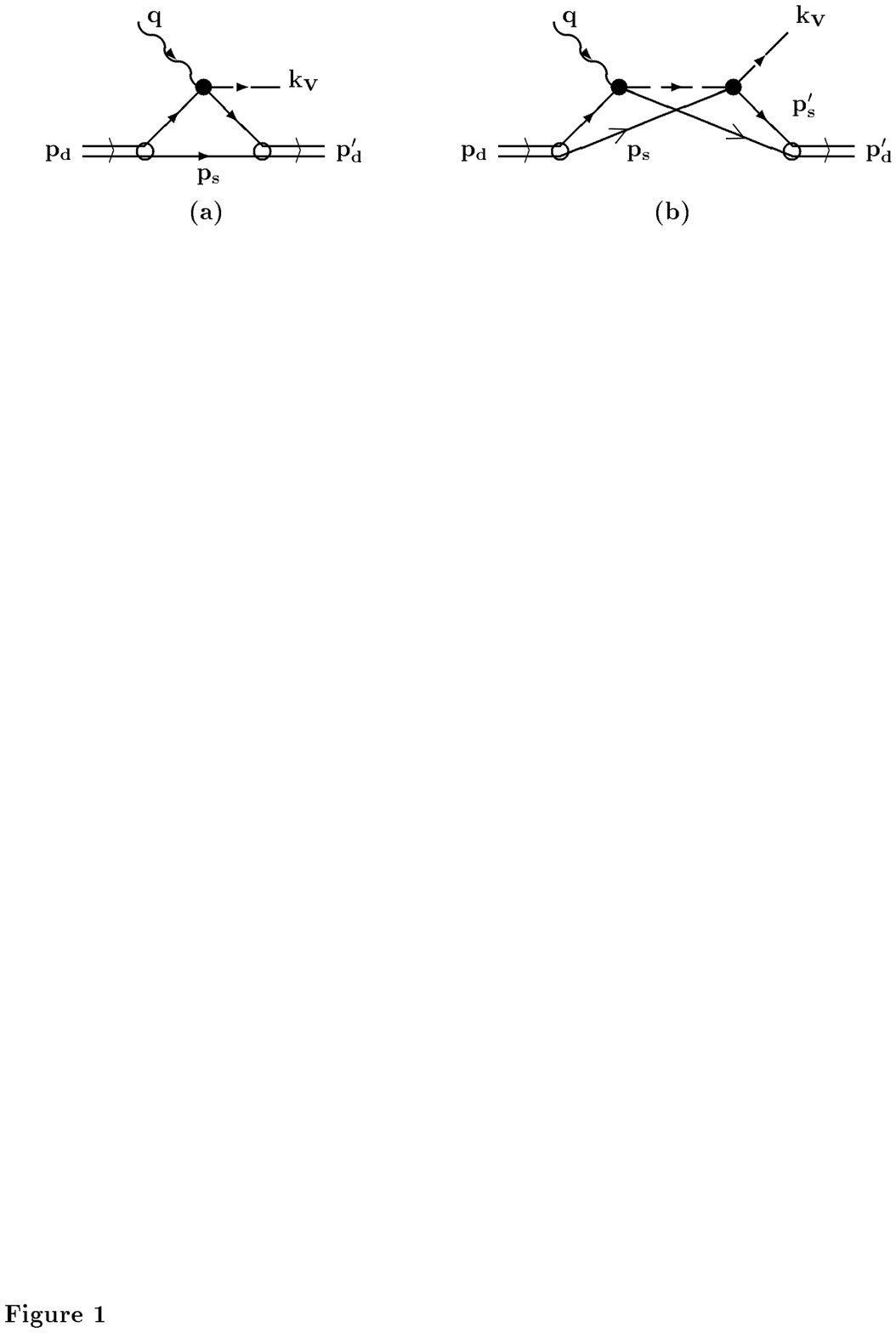}}
\label{Fig.1}
\end{figure}

\vfill
%\noindent {\bf Figure 1}: \ 
{Single scattering  (a) 
and  double scattering contribution  (b) to 
exclusive vector meson production in photon-deuteron 
collisions.}

\newpage
\begin{figure}[t]
\centerline{\epsffile{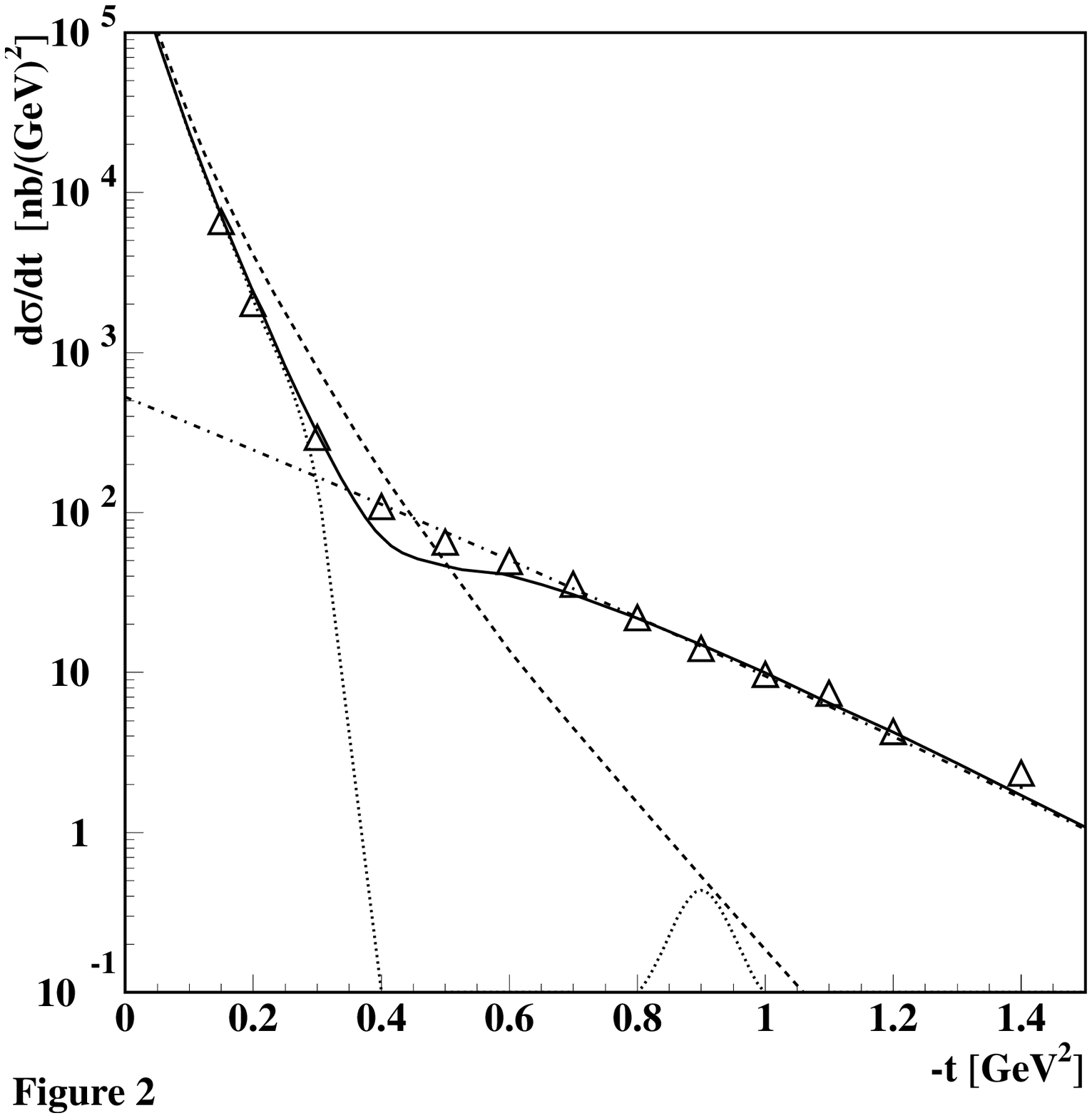}}
\label{Fig.2}
\end{figure}

\vfill

%\noindent {\bf Figure 2}: \ 
{The cross section
$d\sigma_{\gamma d\rightarrow \rho d}/dt$ for photoproduction 
of $\rho$-mesons from unpolarized deuterium.
The full curve shows the result of our calculation within 
vector meson dominance. 
The dashed, dotted and dash-dotted lines account for 
the Born, interference, and double scattering contribution, 
respectively.
The experimental data are from Ref.\cite{Overman,Anderson}
taken at  a photon energy $\nu=12 \,GeV$.}

\newpage
\begin{figure}[t]
\centerline{\epsffile{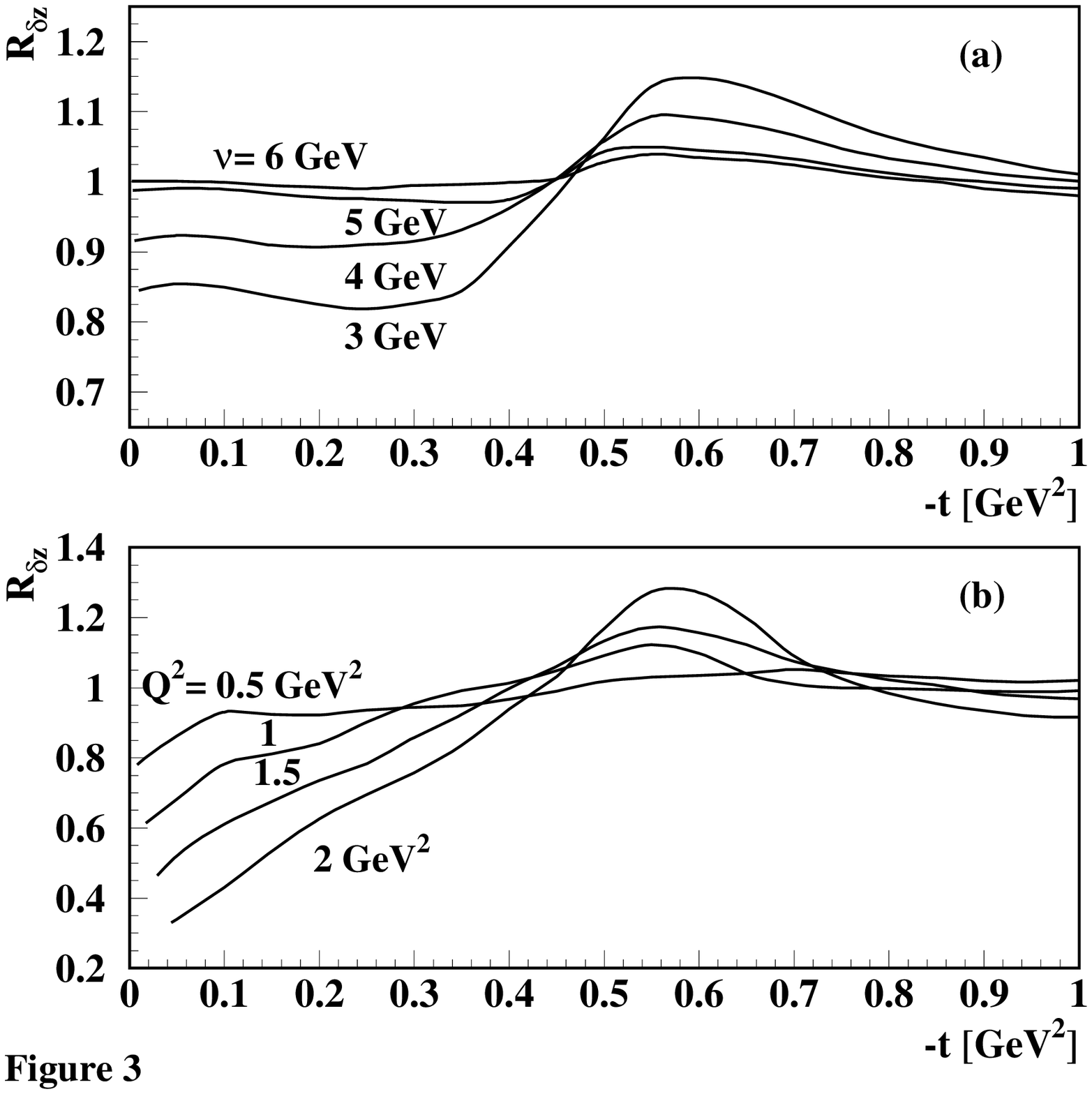}}
\label{Fig.3}
\end{figure}

\vfill
%\noindent{\bf Figure 3}: \ 
{The cross section ratio $R_{\delta_z}$ from (\ref{Rl}) plotted 
against $t$ 
for photoproduction (a) and leptoproduction (b) at 
$\nu=6\,GeV$.}

\newpage
\begin{figure}[t]
\centerline{\epsffile{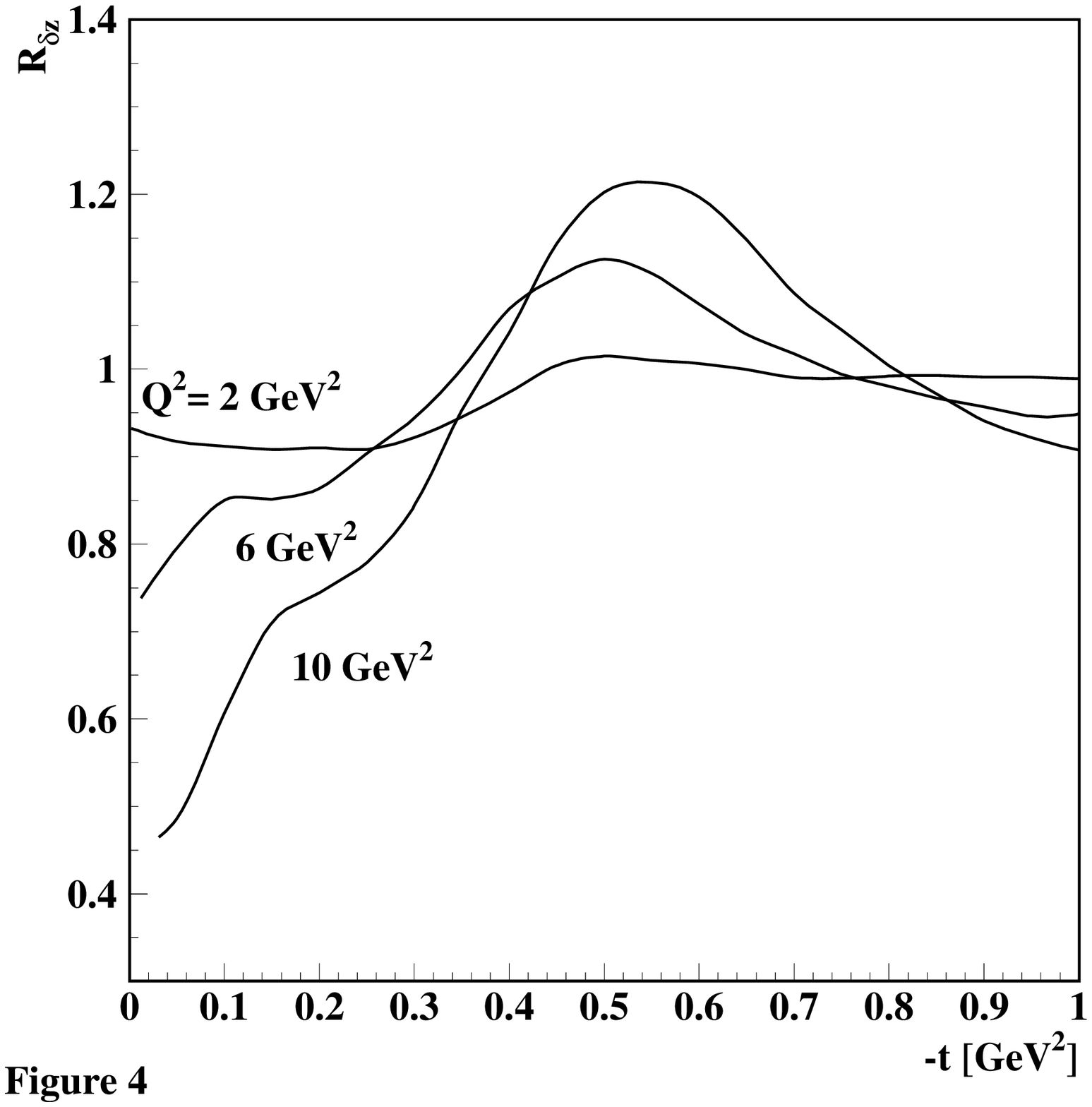}}
\label{Fig.4}
\end{figure}

\vfill
%\noindent{\bf Figure 4}: \ 
{The cross section ratio $R_{\delta_z}$ from (\ref{Rl}) 
for $\nu = 30\,GeV$ and various values of $Q^2$.}

\newpage
\begin{figure}[t]
\centerline{\epsffile{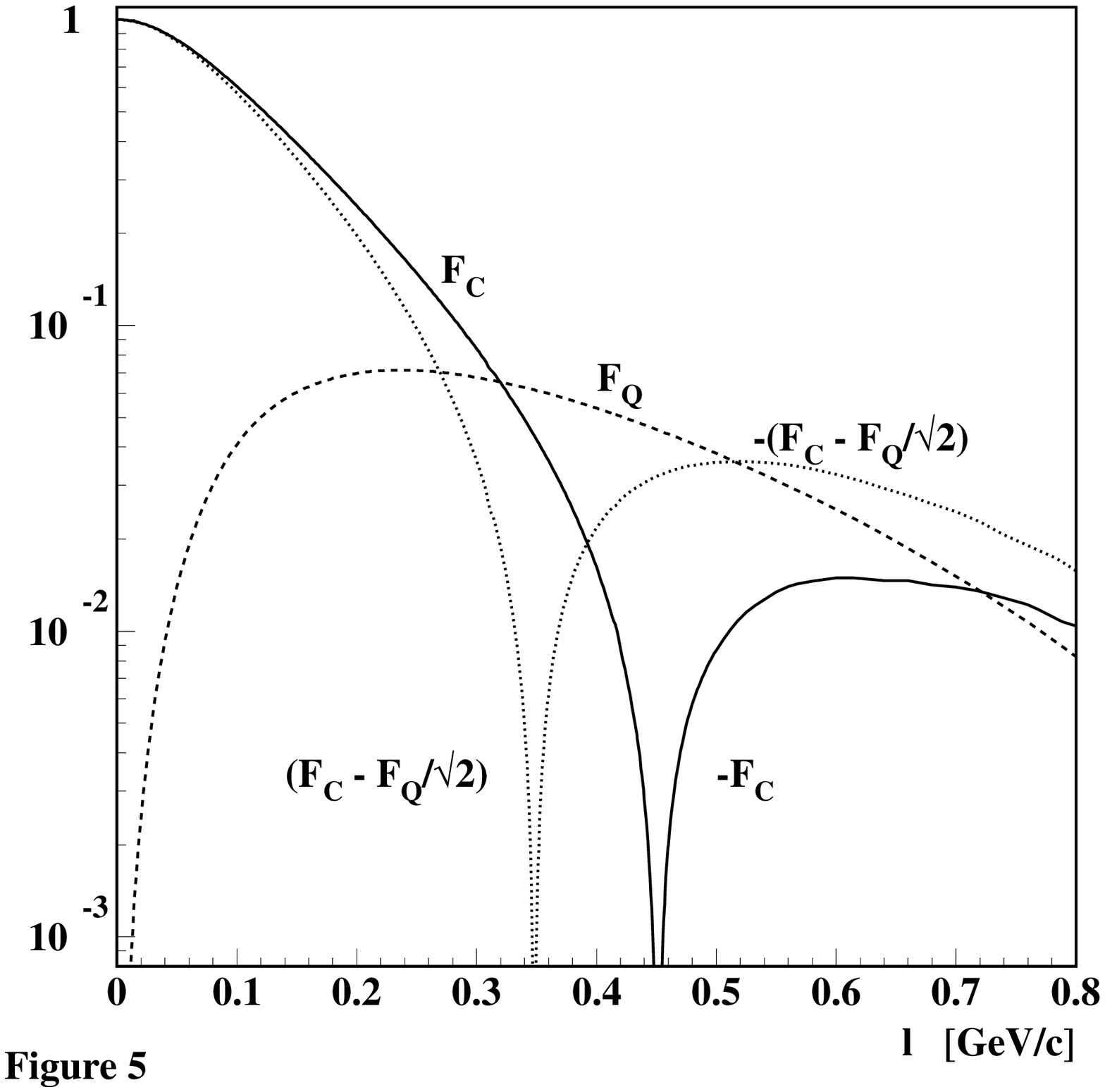}}
\label{Fig.5}
\end{figure}

\vfill

%\noindent{\bf Figure 5}: \ 
{The momentum dependence of the deuteron form factor 
(\ref{eq:form_factors}) for the Paris potential \cite{LaLoRi80}. 
The charge, $F_C$, and quadrupole form factor, $F_Q$, 
are shown as well as the combination $F_C - F_Q/\sqrt{2}$. }

\newpage
\begin{figure}[t]
\centerline{\epsffile{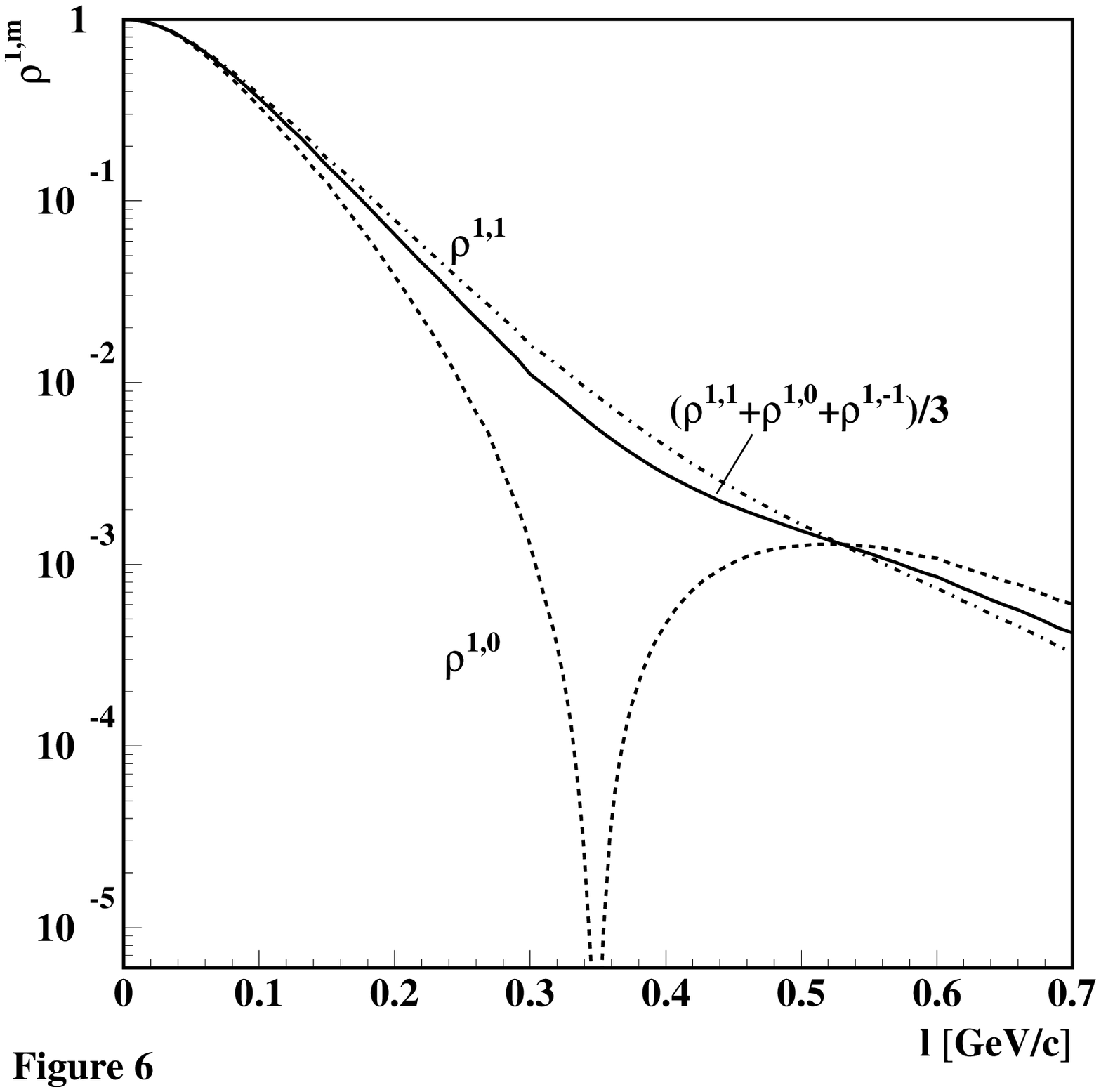}}
\label{Fig.6}
\end{figure}

\vfill

%\noindent{\bf Figure 6 }: \ 
{The deuteron density matrix $\rho^{1,m}$ from 
(\ref{eq:density_matrix}) for a spin quantization 
axis parallel to the photon momentum $\vek q$. 
The dashed and dash-dotted lines correspond  to spin projections 
$m=0$ and $m=1$, respectively.
The density matrix for an unpolarized  deuteron is shown by 
the solid curve.
% while the dotted curve represents the 
%density matrix for tensor polarization (\ref{eq:tensor_pol}).} 

\vfill

\begin{figure}[t]
\centerline{\epsffile{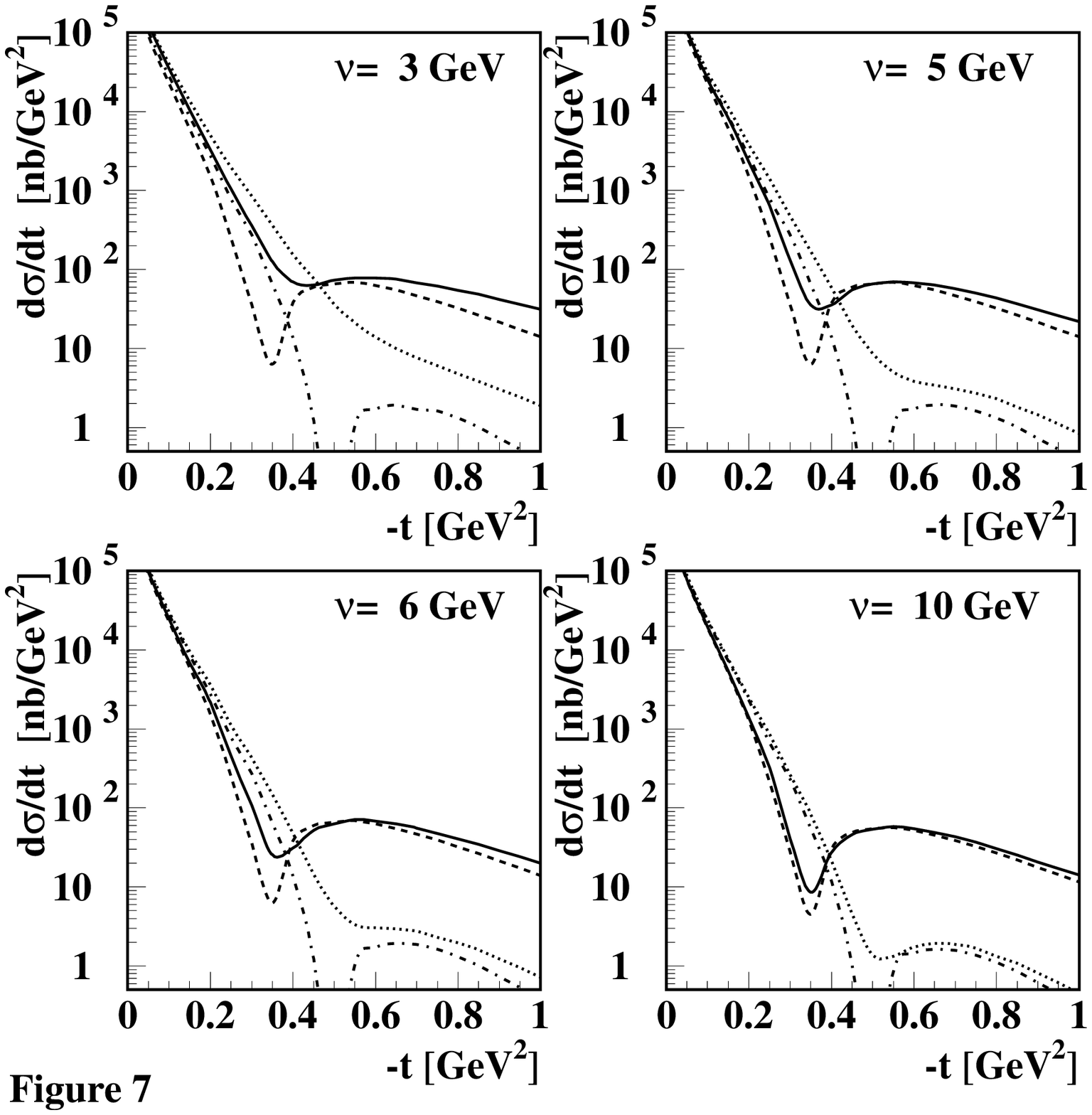}}
\label{Fig.7}
\end{figure}

%\noindent{\bf Figure 7 }: \ 
{The differential cross section 
$d\sigma^{10}_{\gamma d\rightarrow \rho d}/dt$ for the photoproduction 
of $\rho$-mesons from polarized deuterium for different photon energies.
The target polarization is fixed at $m=0$ with respect to 
the photon momentum ($s=1$). 
The solid curves show  the complete vector meson dominance result. 
The dotted curves represent the Born contribution only.
The dashed and dash-dotted curves show the full and the Born 
cross section for infinite longitudinal interaction lengths 
$\delta_z^{(a)}, \delta_z^{(b)} \rightarrow \infty$.
}

\newpage
\begin{figure}
\centerline{\epsffile{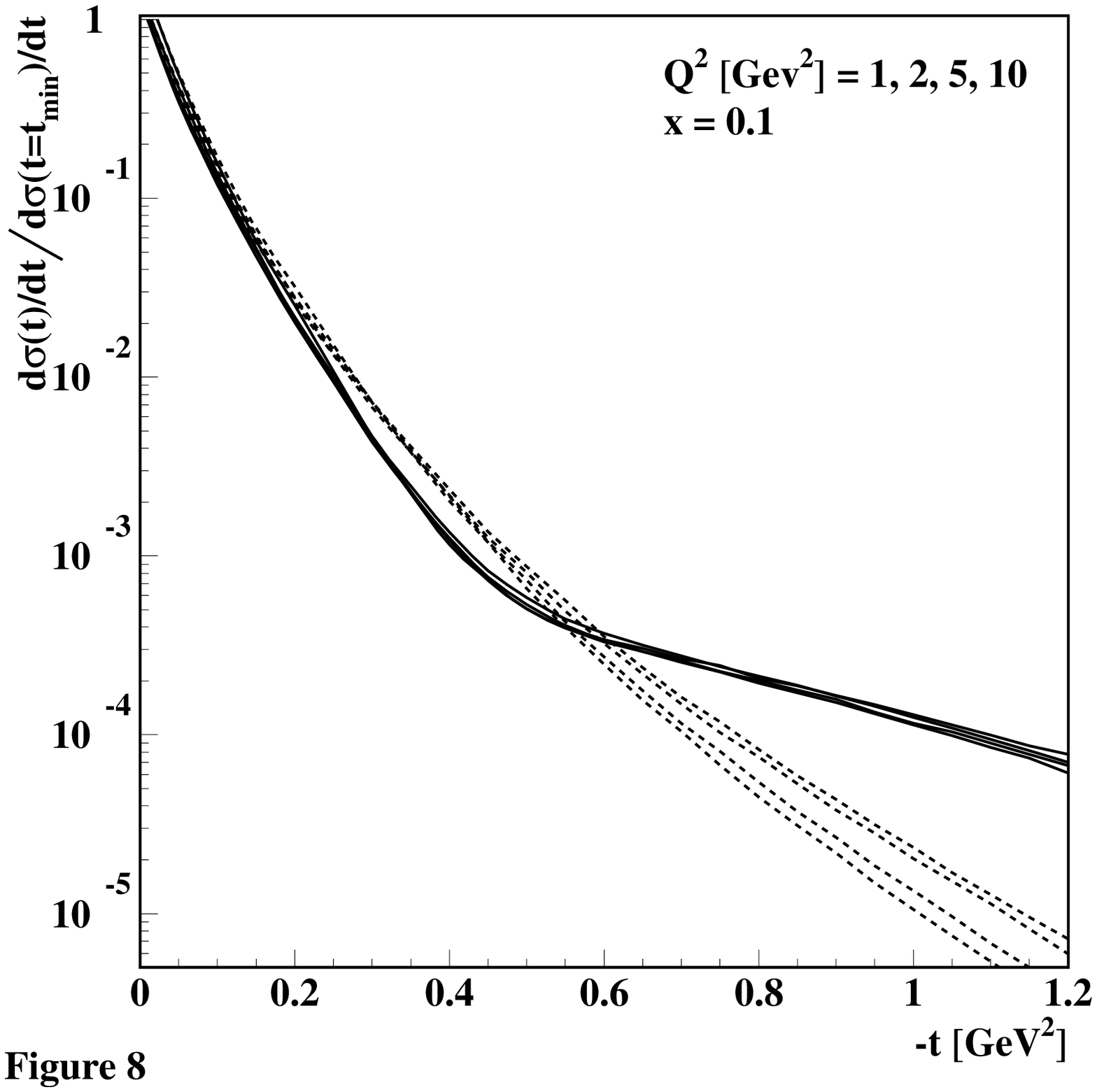}}
\label{Fig.8}
\end{figure}

\vfill

%\noindent{\bf Figure 8}: \ 
{The cross section ratio $R$ from (\ref{R_Q2}) for $x=0.1$ 
and various  values of $Q^2$.
The solid lines represent the complete result of vector meson dominance. 
The dashed curves show  the Born contribution.
}

\newpage
\begin{figure}[t]
\centerline{\epsffile{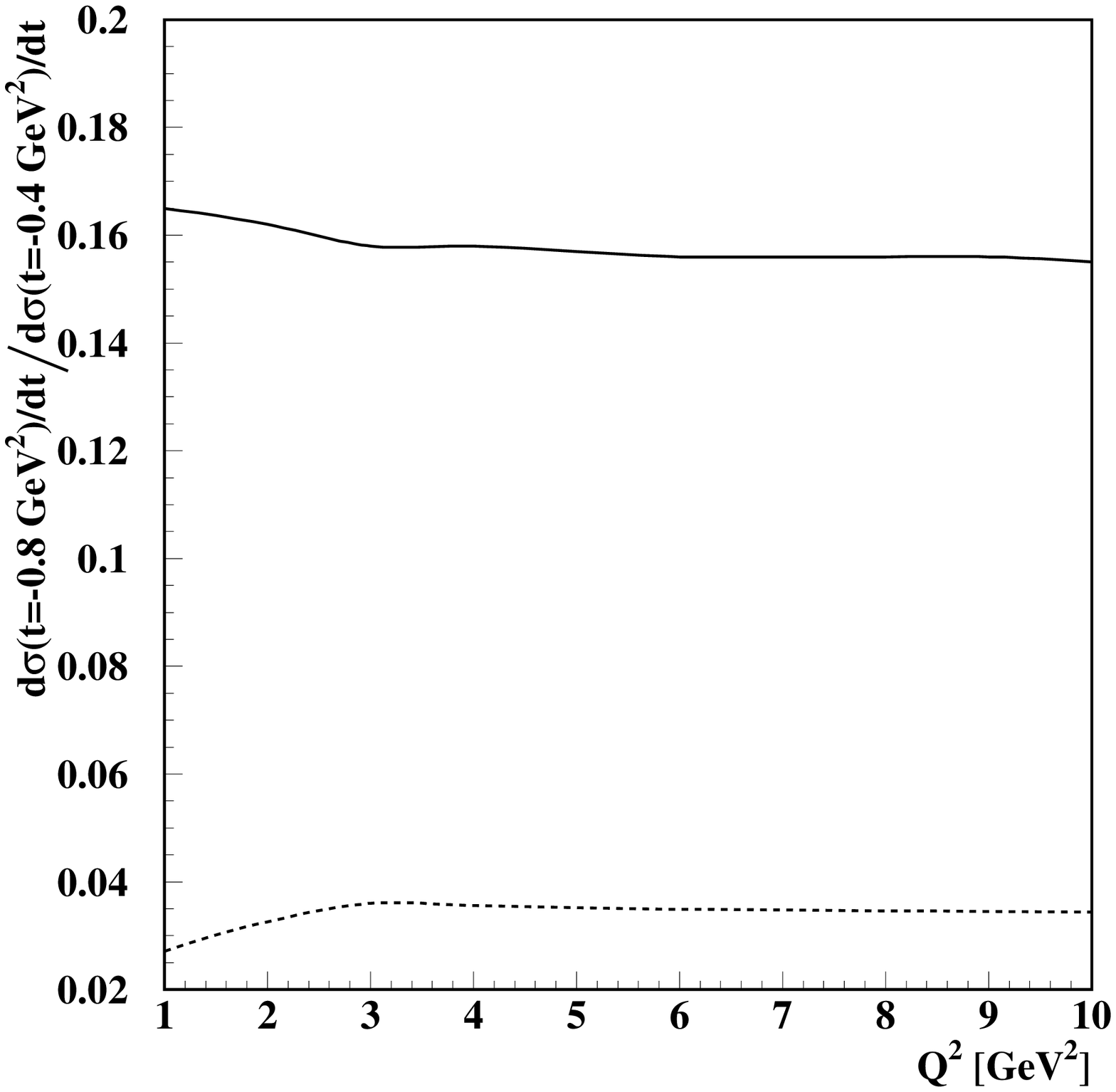}}
\label{Fig.9}
\end{figure}

\vfill

%\noindent{\bf Figure 9}: \ 
{The $Q^2$-dependence of the ratio $R(t=-0.8\,GeV^2)/R(t=-0.4\,GeV^2)$ 
from (\ref{R_Q2}).
The solid line represents the complete vector meson dominance 
calculation. The dashed curve  accounts for  the Born contribution 
only.
}

\begin{figure}[t]
\centerline{\epsffile{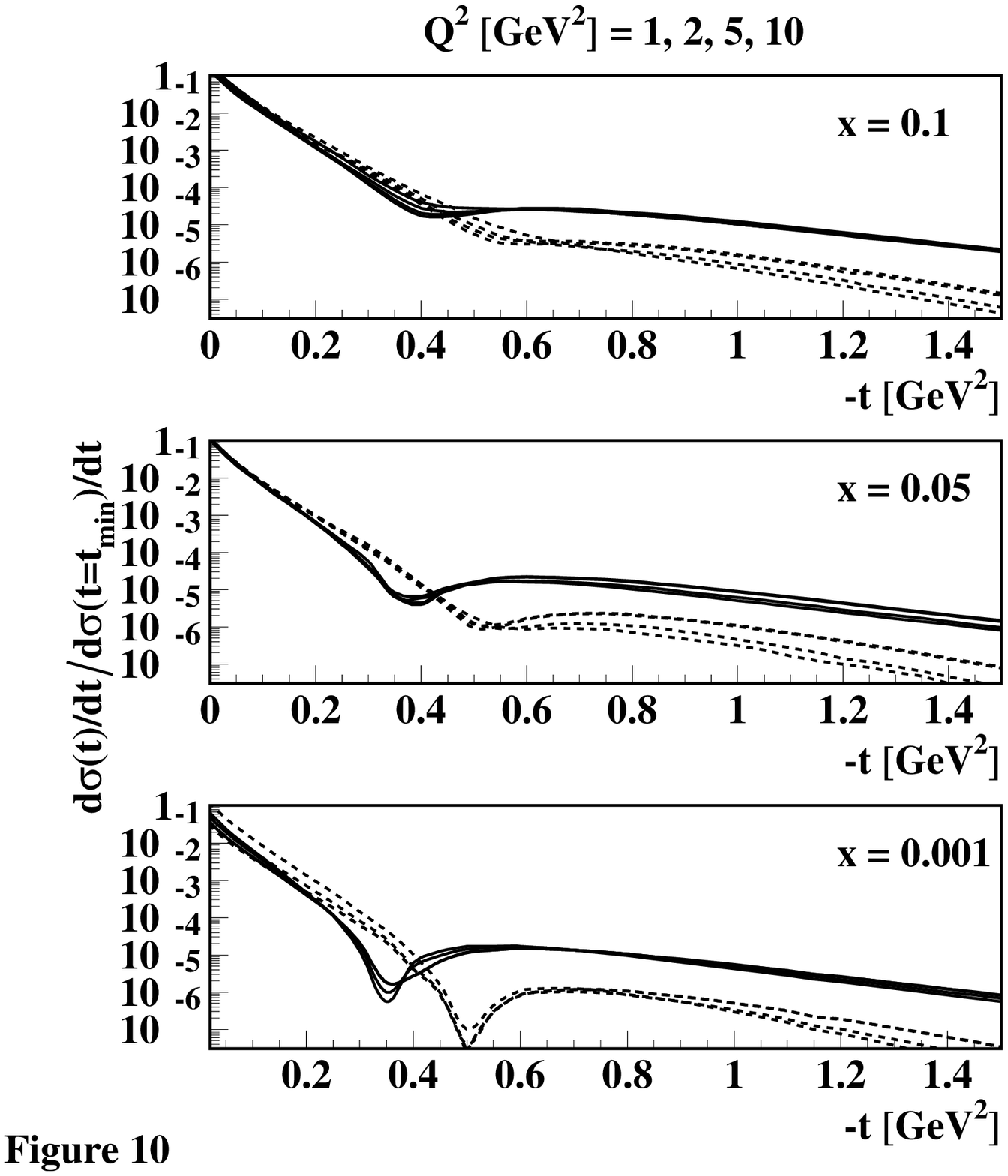}}
\label{Fig.10}
\end{figure}

\vfill

%\noindent{\bf Figure 10}: \
{The cross section ratio $R^{1,0}$ from (\ref{eq:R10}) 
for  $x=0.1, 0.05, 0.01$, and various values of $Q^2$.
The solid lines represent the complete vector meson dominance 
calculation. The dashed curves show  the Born contribution.
}

\vfill

\begin{figure}[t]
\centerline{\epsffile{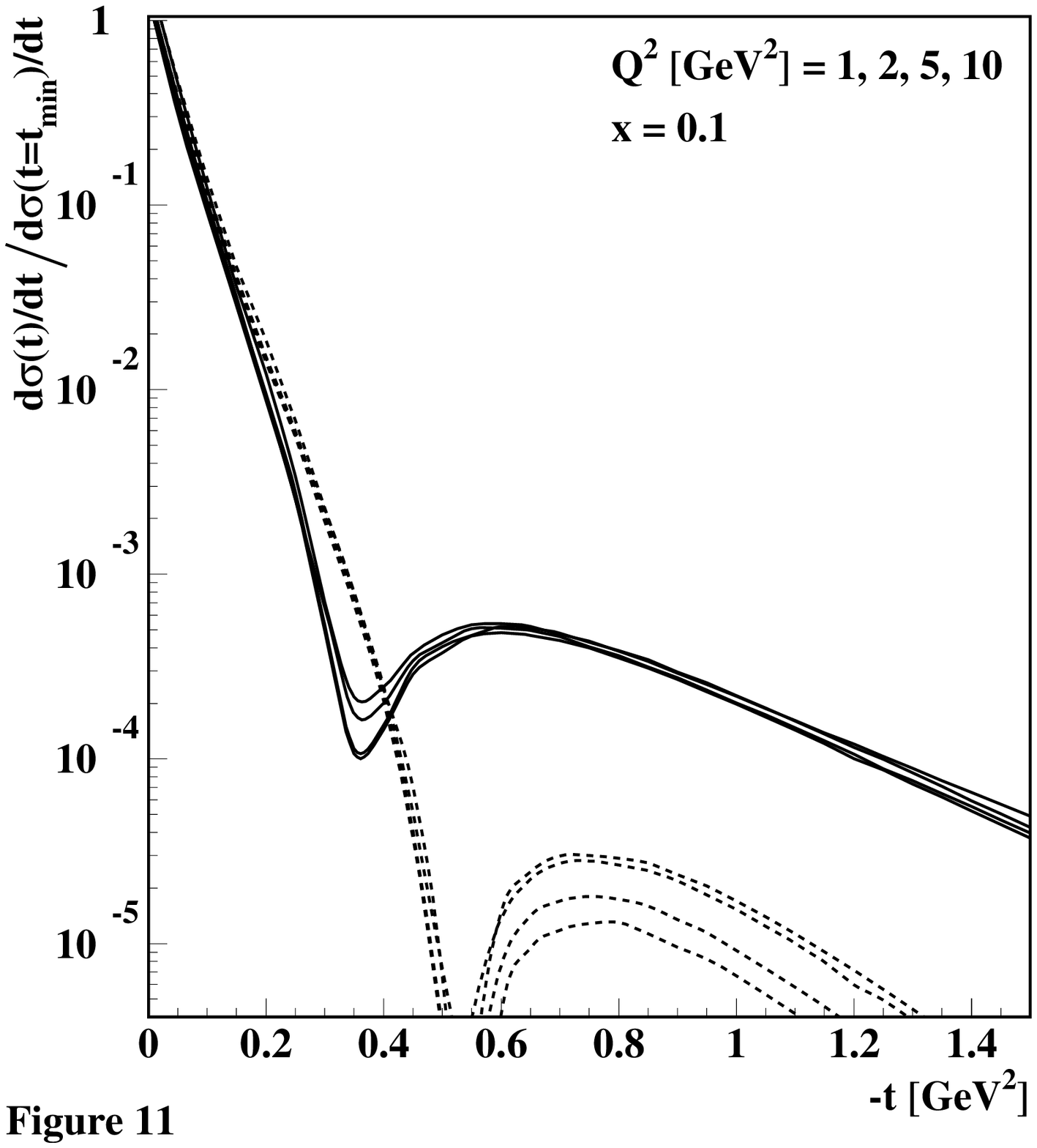}}
\label{Fig.11}
\end{figure}

\vfill

%\noindent{\bf Figure 11}: \ 
{The cross section ratio $R^{2,0}$ (\ref{deny})  
for  $x=0.1$ and various values of $Q^2$.
The solid lines represent  the complete  vector meson dominance 
calculation. The dashed curves show  the Born contribution.
}

\newpage

\begin{figure}
\centerline{\epsffile{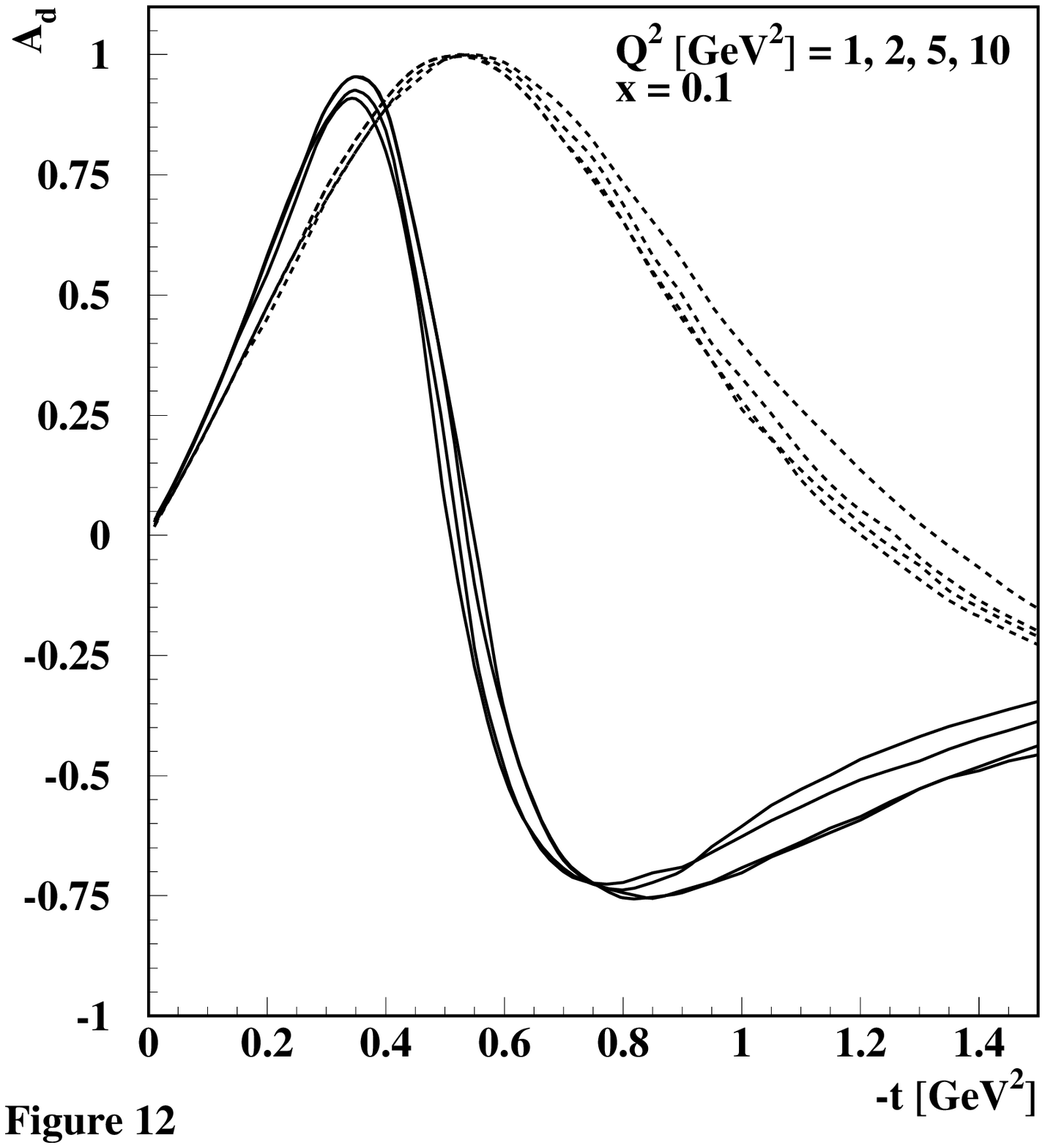}}
\label{Fig.12}
\end{figure}

\vfill

%\noindent{\bf Figure 12}: \ 
{The tensor polarization asymmetry $A_d$ from (\ref{eq:tensor_pol}) 
for $x=0.1$ and various values of $Q^2$. 
The spin quantization axis is chosen parallel to 
$\vek \kappa \sim \vek q\times \vek l$. 
The solid lines represent the complete vector meson dominance 
calculation. The dashed curves show  the Born contribution.
}

\end{document}